\documentclass[nofootinbib,notitlepage,onecolumn,secnumarabic,amssymb, nobibnotes, aps, prd]{revtex4-1}
\usepackage{graphicx,epsfig}
\pdfoutput=1
\usepackage{upgreek}
\usepackage{amsmath}
\usepackage{amssymb}
\usepackage{cancel}
\usepackage{float}
\usepackage{bm}
\usepackage{multirow}
\usepackage{amsfonts}
\usepackage{xcolor,cancel}
\usepackage{hyperref}
\usepackage{epstopdf}
\usepackage{natbib}
\usepackage{hyperref}
\usepackage{enumerate}
\usepackage{afterpage}
\usepackage{array}
\usepackage{placeins}
\usepackage{calc}

\newcommand{\HH}{\mathcal{H}}
\newcommand{\n}{{\bm {n}}}
\newcommand{\p}{\partial}
\newcommand{\Q}{\mathcal{Q}}
\newcommand{\fnl}{{f}_\text{NL}}

\def\k {\bm{k}}
\def\x {\bm{x}}
\def\d {\mbox{d}}

\newcommand{\<}{\langle}
\renewcommand{\>}{\rangle}

\def\be{\begin{equation}}
\def\ee{\end{equation}}
\def\bea{\begin{eqnarray}}
\def\eea{\end{eqnarray}}

\begin{document}

\title{\large Imprints of local lightcone projection effects on the galaxy bispectrum. {II}}
\author{\large  Sheean Jolicoeur$^a$, Obinna Umeh$^a$,  Roy Maartens$^{a,b}$ and Chris Clarkson$^{a,c,d}$ \\~\\~\\
\emph{\normalsize $^a$Department of Physics \& Astronomy, University of the Western Cape,
Cape Town 7535, South Africa \\
$^b$Institute of Cosmology \& Gravitation, University of Portsmouth, Portsmouth PO1 3FX, United Kingdom\\
$^c$School of Physics \& Astronomy, Queen Mary University of London, London E1 4NS, United Kingdom\\
$^d$Department of Mathematics \& Applied Mathematics, University of Cape Town, Cape Town 7701, South Africa}~\\~\\}

\date{\today}

\begin{abstract}
~\\

\large General relativistic imprints on the galaxy bispectrum arise from  observational (or projection) effects. The lightcone projection effects include  local contributions from Doppler and gravitational potential terms, as well as  lensing and other integrated contributions. We recently presented for the first time, the correction to the  galaxy bispectrum  from all local lightcone projection effects up to second order in perturbations. Here we provide the details underlying this correction, together with further results and illustrations. For moderately squeezed shapes, the correction to the Newtonian prediction is {$\sim 30\%$ on equality scales at $z\sim 1$}. We generalise our recent results to include the contribution, up to second order, of  magnification bias (which affects some of the local terms) and evolution bias.\par

\end{abstract}

\maketitle
\tableofcontents
\clearpage
\normalsize

\section{Introduction}

The galaxy power spectrum has been central to the cosmological constraints extracted from galaxy surveys up to now. 
For an accurate comparison of observations to theory, observational projection effects on the galaxy power spectrum must be taken into account. The main projection effect comes from redshift-space distortions (RSD) {\cite{Jackson:2008yv,Sargent:1977ApJ...212L...3S,Kaiser:1987qv}}, which must be included in the analysis of the power spectrum. But it is not only accuracy that is gained -- there is additional information to be extracted from the RSD themselves. 

In addition to RSD, the galaxy power spectrum is also affected by lensing magnification {\cite{Moessner:1997vm,Hui:2007cu,Hui:2007tm}}. In the analysis of current surveys, the lensing contribution to galaxy number counts is typically not included in the power spectrum. For future surveys, which will probe higher redshifts,  this lensing projection effect will need to be included in the galaxy power spectrum for an accurate theoretical analysis -- and, as with RSD, the lensing itself will deliver additional information \cite{Alonso:2015uua,Montanari:2015rga,Dizgah:2016bgm}. 

Lensing convergence contributes a general relativistic (GR) projection effect, which is a correction to the Newtonian (overdensity + RSD) galaxy power spectrum. There are further GR projection effects which {modify} the galaxy power spectrum on ultra-large scales ($ H_0\lesssim k\lesssim k_{\rm eq}$) \cite{Yoo:2010ni,Challinor:2011bk,Bonvin:2011bg}. These include Doppler, Sachs-Wolfe, integrated Sachs-Wolfe and time-delay terms. 
As in the case of RSD and lensing, these terms need to be incorporated for accuracy, and they also contain extra information. 

The ultra-large scale GR corrections have a qualitatively similar effect on the galaxy power spectrum to primordial non-Gaussianity (PNG), which {also modifies the power spectrum} on ultra-large scales via scale-dependent galaxy bias. The GR corrections must therefore be taken into account when super-equality scales are probed to measure or constrain the PNG parameter $\fnl$ \cite{Bruni:2011ta,Baldauf:2011bh,Jeong:2011as,Camera:2014bwa,Camera:2014sba}.

The galaxy bispectrum can provide additional information, partly independent of the power spectrum {\cite{Sefusatti:2006pa,Sefusatti:2007ih}}.
{The effects on the bispectrum from RSD have been computed in \cite{Verde:1998zr,Scoccimarro:1999ed} and from lensing in \cite{Schmidt:2008mb}}.
Recently, the galaxy bispectrum has been used to detect the RSD and baryon acoustic oscillation (BAO) features in the BOSS survey, and to give independent measurements of growth rates and distances \cite{Gil-Marin:2016wya,Slepian:2016kfz}. 

As in the case of the galaxy power spectrum, we need to take account of the observational lightcone effects in the galaxy bispectrum which distort the information on the underlying dark matter distribution, but which also provide new information. These projection effects are the same as for the power spectrum -- with one major difference: for the bispectrum, we require the projection effects up to at least second order in perturbations. 

Next-generation galaxy surveys will enable increasingly accurate measurements of the galaxy bispectrum, out to higher redshifts and across larger sky areas. Recent forecasts, using a Newtonian model with RSD but no GR projection effects, indicate that the bispectrum can considerably enhance the constraining power of future surveys \cite{Tellarini:2016sgp} -- especially for probing the initial conditions of the Universe via PNG. In order to fully exploit the improved precision from upcoming surveys, we need theoretical accuracy that matches and moves beyond observational precision.  One important part of this theoretical requirement is to include all the GR projection effects in modelling the galaxy bispectrum. 

The GR lightcone effects on the galaxy angular bispectrum from lensing convergence were computed on intermediate scales in \cite{DiDio:2015bua}, neglecting the other, ultra-large scale, GR corrections to the galaxy overdensity. 
Another partial result was given in \cite{Kehagias:2015tda},  using a separate-universe approximation to compute the galaxy angular bispectrum with all GR lightcone effects in the squeezed limit. 

We recently provided a further partial result, valid for all triangle shapes, by computing all the local GR projection corrections to the galaxy bispectrum, including all second order terms and couplings \cite{Umeh:2016nuh}.  
Crucial to our result is the  expression for the observed galaxy number counts on the past lightcone, up to second order. This is given in the  most general case by \cite{Bertacca:2014hwa} (see also \cite{Bertacca:2014dra,Bertacca:2014wga,Yoo:2014sfa,DiDio:2014lka}). 
Our work is complementary to the subsequent work by \cite{DiDio:2016gpd}, who include lensing and terms of order $(\HH/k)[\delta^{(1)}]^2$, but neglect all other GR effects on ultra-large scales.

Here we provide details of the derivation of the results given in  \cite{Umeh:2016nuh}, with additional illustrations, and we generalise some of those results. In particular, we include the magnification bias (which also contributes to local terms in the number counts) and the evolution bias. In \cite{Umeh:2016nuh},  both of these were set to zero.

{We focus on large enough scales that perturbation theory is accurate, and we make the following assumptions:
\begin{itemize}
\item
A Gaussian primordial curvature perturbation.
\item
A simple local-in-mass-density model of galaxy bias, as in \cite{Sefusatti:2006pa,Sefusatti:2007ih} (schematically, $\delta_g \sim b_1\delta_m+b_2\delta_m^2/2$). However, we take care to ensure that the definition of bias is gauge-independent and applies on ultra-large scales.
\item
For simplicity, we use standard Newtonian results to evaluate 
the second-order velocity potential $v^{(2)}$ and metric potentials $\Phi^{(2)},\Psi^{(2)}$, which contribute to the projection effects.
\item
We neglect the second-order effect of the radiation era on 
 initial conditions for sub-equality modes \cite{Tram:2016cpy}.
\item
We compute the galaxy bispectrum at fixed redshift and in Fourier space, and we use the plane-parallel approximation. Consequently, the following are not included in our approach: wide-angle correlations, radial correlations, lensing and other integrated contributions. 
 
\end{itemize}

At second order in GR, scalar perturbations generate secondary vector and tensor modes  \cite{Mollerach:1997up,Matarrese:1997ay}. These modes also enter the projection effects in the observed galaxy number density contrast at second order \cite{Bertacca:2014hwa,Bertacca:2014dra,Bertacca:2014wga,Yoo:2014sfa,DiDio:2014lka}. As shown by  \cite{Lu:2008ju,Bruni:2013mua} for vector modes and \cite{Ananda:2006af,Baumann:2007zm,Jeong:2012nu} for tensor modes, the power in the secondary vector and tensor modes is 
 much smaller than the scalar power at second order, so we neglect
the vector and tensor contributions. }

We adopt a standard concordance model, with parameters given by the latest Planck best-fit values \cite{Ade:2015xua}; in particular,  $h =H_0/(100\,{\rm km \, s}^{-1}{\rm Mpc}^{-1})= 0.678$ and  $\Omega_{m0} =1-\Omega_{\Lambda 0}= 0.308$.

\newpage
\section{Galaxy number counts in general relativity}\label{sec:reviewdaniele}

The observer looks down the past lightcone and  counts $\d N$ galaxies, above a threshold luminosity $L$, within a redshift interval $\d z$ about the observed redshift $z$, and within a solid angle element $\d\Omega_o$ about the observed direction $\bm n$, where \cite{Challinor:2011bk,Jeong:2011as,Bertacca:2014hwa,Alonso:2015uua}
\begin{equation}\label{eq:Numbercount}
{\d N}(z,\bm n,>\ln L) = \mathcal{N}(z,{\bm{n}}, >\ln L)\, D^2_A(z,{\bm{n}}) \,k_{\mu} u^\mu\, \frac{\d \lambda}{\d z}\, {\d z\, \d \Omega_o}\,.
\end{equation}
Here $D_A$ is the angular diameter distance, $u^\mu$ is the 4-velocity of the source,  $k^\mu = \d x^\mu/\d \lambda$  is the geodesic photon 4-momentum, and $\mathcal{N}$ is the flux-limited  number density of sources:
\begin{eqnarray}\label{eq:fluxlimit}
\mathcal{N}(z,{\bm{n}}, >\ln L) = \int_{\ln L}^{\infty} \d\ln{\tilde L} \,n_g(z,{\bm{n}},\ln \tilde L)\,.
\end{eqnarray}
In the integrand, $n_g$ is the proper number density of sources, 
and only sources with luminosity above the detection threshold are counted by the observer.

The fractional perturbation $\Delta_g$ of the observed number counts is defined by
\begin{eqnarray}
  \frac{\d N(z,\bm n,{>\ln L})}{\d z \d \Omega_o}& =& \frac{\chi^2(z)}{(1+z)^4 \HH(z)}\bar{\mathcal{N}}(z,{>\ln L}) \big[1+ {\Delta_g}(z,{\bm{n}},{>\ln L})\big], \label{dgdef}
  \end{eqnarray}
 where 
 $\HH(\eta)=a'(\eta)/a(\eta)$ is the conformal Hubble rate,  the comoving line-of-sight distance is given by $\d\chi=\d z/[(1+ z)\HH( z)]$, and $\bar{\mathcal{N}}$ is  the background  magnitude-limited number density. Henceforth, we  suppress the dependence of $\Delta_g$ on ${\ln L}$ to reduce clutter. 
 We expand ${\Delta_g}$  up to second order in perturbation theory:
  \begin{eqnarray}\label{eq:perturbation}
 {\Delta_g}(z,{\bm{n}}) &=&  \Delta_{g}^{{{({1})}}}(z,{\bm{n}}) + \frac{1}{2}\left[ \Delta_{g}^{{{({2})}}} (z,{\bm{n}})  - \big\< \Delta_{g}^{{{({2})}}}(z,{\bm{n}})\big\>\right] ,
     \end{eqnarray}
 where we subtract off the average of ${\Delta_g^{{{({2})}}}}$  in order to ensure that $\<{\Delta_g}\> = 0$.     
For later convenience, we split the observed number density contrast into Newtonian  and GR parts:
\begin{eqnarray}
\Delta_g^{(r)} = \Delta^{(r)}_{g{\rm N}} + \Delta^{(r)}_{g{\rm GR}},\qquad r=1,2\,.
\end{eqnarray}

We only consider the bispectrum at fixed redshift, so that all correlations are in the same redshift bin. {There are integrated GR contributions to $\Delta_g^{(1)}$, from weak lensing convergence and also from integrated Sachs-Wolfe and time-delay terms, and we neglect these terms. 
At second order, there are many more terms with line-of-sight integrated contributions, and we neglect all such terms. Specifically, we neglect the integrated contributions in   \cite{Bertacca:2014hwa}, which gives the fully general  $\Delta_g^{(1)}$ and $\Delta_g^{(2)}$ in Poisson gauge.\footnote{We also neglect all terms at the observer, {which do not contribute to the bispectrum}.}} A complete treatment would include the integrated terms, with all cross-bin correlations. This far more complicated analysis is left for future work.

An important point to note is that the GR weak lensing convergence consists not only of the standard integrated term, but also includes local (non-integrated) terms \cite{Bonvin:2008ni}. This means that the magnification bias will still enter the bispectrum, even if we neglect all integrated terms.
The magnification bias is given by the logarithmic slope of the background number density at the threshold luminosity:
\begin{eqnarray}
\label{Q}
\mathcal{Q}(a,\bar L) 
=-\frac{ \partial \ln\big[a^3 \bar{\mathcal{N}}( a , {>}\bar L)\big]}{\partial\ln \bar L}.
\end{eqnarray}
We have used the comoving number density in the definition above since it arises also in the definition of the evolution bias:
\be\label{be}
b_e(a,\bar L) =\frac{\partial \ln\big[a^3 \bar{\mathcal{N}}( a , {>}\bar L)\big]}{\partial \ln  a}.
\ee
This quantity describes the deviation of the background number density of sources from the idealised case of $a^3\bar{\mathcal{N}}= \bar{\mathcal{N}}_0$.

{Radial and transverse derivatives are defined as} 
\be \label{rtder}
\partial_\|=n^i\partial_i,\qquad \qquad  \partial_{\perp i}= \partial_i-n_i \partial_\|,
\ee    
 the derivative down rays of the past lightcone is 
\begin{equation}\label{chi}
{\d\over\d\chi}=-{\d\over\d\eta}=-\partial_{\eta} + \partial_{\parallel} \,,
\end{equation}
and the screen space projected Laplacian is 
\begin{equation}\label{e19}
\nabla^{2}_{\perp} = \nabla^{2} - \partial_{\parallel}^{2} - \frac{2}{\chi}\partial_{\parallel}\,.
\end{equation}

Since $\Delta_g$ is defined as an observable, it is gauge-independent and we can use any gauge to compute it. In a given gauge, it will be of the form $\Delta_g=\delta_g+$ terms that describe projection effects in that gauge, where $\delta_g=\delta{\cal N}/\bar{\cal N}=\delta_g^{{{({1})}}}+{1\over2} \delta_g^{{{({2})}}}$ is the galaxy number density contrast in the chosen gauge.
We choose the Poisson gauge since it is convenient for splitting into Newtonian and GR parts. Neglecting the vector and tensor modes, the metric and the peculiar velocity of galaxies (equal to the dark matter velocity on the scales of interest) are given by
\bea
a^{-2}\d s^2 &=& -\left[1+2\Phi^{{{({1})}}}+\Phi^{{{({2})}}}\right]\d\eta^2+ \left[1-2\Phi^{{{({1})}}}-\Psi^{{{({2})}}}\right]\d{\bm x}^2,\\
v^i &=& \partial^i\Big[ v^{{{({1})}}}+{1\over2}v^{{{({2})}}} \Big].
\eea
The {observed  comoving coordinates \cite{Bertacca:2014dra}} of a galaxy are $\bm x=\chi(z)\bm n=[\eta_0-\eta(z)]\bm n$.
 We have assumed that anisotropic stress vanishes {at first order}, which implies  $\Psi^{{{({1})}}} =\Phi^{{{({1})}}}$ in GR. 

We will also use the comoving-synchronous (C) overdensities of matter and galaxy counts $\delta_{m{\rm C} }, \delta_{g{\rm C} }$. The first-order Poisson and continuity equations are then
\be \label{pc}
\nabla^2\Phi^{{{({1})}}}= \frac{3}{2}\Omega_m {\HH^2}\,\delta_{m{\rm C} }^{{{({1})}}},\qquad
\delta_{m{\rm C} }^{{{({1})\prime}}}=-\nabla^2v^{{{({1})}}},
\ee
which lead to 
\bea  
\label{phi}
\Phi^{{{({1})}}}= -\frac{3}{2}\Omega_m \frac{\HH^2}{k^2}\,\delta_{m{\rm C} }^{{{({1})}}} ~~&&\mbox{where}~~  \Phi^{{{({1})}}}(a,\bm{k}) = {D(a) \over a}\, \Phi^{{{({1})}}}(1,\bm{k}), \\
\label{vm}
\HH v^{{{({1})}}} = f\frac{\HH^2}{k^2} \,\delta_{m{\rm C} }^{{{({1})}}}~~&&\mbox{where}~~ f={\d\ln D \over \d\ln a} ~~\mbox{and}~~ \delta_{m{\rm C} }^{{{({1})}}}(a,\k)=D(a)\, \delta_{m{\rm C} }^{{{({1})}}}(1,\k).
\eea

{\subsection{Local model of galaxy bias on ultra-large scales}}

We start by considering the Poisson-gauge number density contrast $\delta_g^{{{({1})}}}$ at linear order, which is related to the dark matter density contrast  $\delta^{{{({1})}}}_m$ via the galaxy bias. We need to ensure that the definition of scale-independent galaxy bias is gauge-independent  and valid on ultra-large scales. As explained in detail in \cite{Challinor:2011bk,Bruni:2011ta,Jeong:2011as}, the physical definition of scale-independent bias is in the matter rest-frame, which coincides with the galaxy rest-frame (on large scales there is no velocity bias). The matter rest-frame corresponds to the C  gauge, so that the correct definition at first order is (restoring the dependence on $L$):
\be \label{sb}
\delta_{g{\rm C} }^{{{({1})}}}(a,\x,{< \ln L})=b_1(a,{\ln \bar L})\,\delta_{m{\rm C} }^{{{({1})}}}(a,\x).
\ee 
The Poisson-gauge number density contrast is related to the  C-gauge one by \cite{Challinor:2011bk}
\begin{equation}\label{eq:lineargaugetrans}
\delta_{g }^{{{({1})}}}=\delta_{g{\rm C} }^{{{({1})}}}+\left(3- b_e \right)\HH v^{{{({1})}}}
=b_1\delta_{m{\rm C} }^{{{({1})}}}+\left(3- b_e \right)\HH v^{{{({1})}}}.
\end{equation}
The  velocity potential term in \eqref{eq:lineargaugetrans} {ensures gauge-independence of the bias model on ultra-large scales.} This term
is the GR part of $\delta_g^{{{({1})}}}$, since it is suppressed on small scales but grows on ultra-large scales, as shown by \eqref{vm}.

In GR, the Lagrangian frame corresponds to the C gauge \cite{Bertacca:2015mca,Villa:2015ppa}. 
There is no unique Eulerian frame in GR, but a convenient choice is the total-matter (T) gauge. This is related to the C gauge by a purely spatial transformation, so that at first order, the matter and galaxy overdensities are the same \cite{Bertacca:2015mca}:
\be\label{1}
\delta_{m{\rm C}}^{(1)}=\delta_{m{\rm T}}^{(1)}, \qquad   \delta_{g{\rm C}}^{(1)}=\delta_{g{\rm T}}^{(1)}=b_1\delta_{m{\rm T}}^{(1)}.
\ee 
The last equality is the definition of the Eulerian bias parameter at first order. This means that $b_1$ in \eqref{sb} is the Eulerian bias parameter.

{We extend \eqref{sb} to higher order with the simplest possible model of scale-independent bias. This model assumes that galaxy number density contrast is a local function of only the matter density contrast -- the so-called local-in-mass-density model. For a physical definition valid on ultra-large scales, we require that the bias coefficients are scale-independent in the galaxy rest-frame, i.e. in C gauge.
Expanding in powers of the mass density contrast, we have
\be\label{limd}
\delta_{g{\rm C}}=b_1 \delta_{m{\rm C}}+  {1\over 2} b_2\big(\delta_{m {\rm C}}\big)^2 + \cdots,
\ee
where $b_I=b_I(a,\ln L)$.
At first order, this recovers \eqref{sb}. At second order we have:}\footnote{{For convenience, we have omitted the term $-b_2{\big\langle  \big[\delta_{m {\rm C}}^{{{({1})}}}\big]^2\big\rangle}$ on the right of \eqref{gs2}.}}
\be \label{gs2}
\delta_{g{\rm C}}^{{{({2})}}}=b_1 \delta_{m{\rm C}}^{{{({2})}}}+   b_2\big[\delta_{m {\rm C}}^{{{({1})}}}\big]^2.
\ee
 The relation between C- and T-gauge matter overdensities at second order is \cite{Bertacca:2015mca,Villa:2015ppa}
\be\label{15}
\delta_{m{\rm T}}^{(2)}
=\delta_{m{\rm C}}^{(2)}+{2}\big[\partial_i\delta_{m{\rm C}}^{(1)}\big]\nabla^{-2}\partial^i\delta_{m{\rm C}}^{(1)},
\ee
where $-2\nabla^{-2}\partial^i\delta_{mC}^{(1)}$ is a gauge generator. Since the C$\,\to\,$T gauge transformation is purely spatial, \eqref{15} also applies to the galaxy counts:
\be
 \label{15a}
\delta_{g{\rm T}}^{(2)}={\delta_{g{\rm C}}^{(2)}+2\big[\partial_i\delta_{g{\rm C}}^{(1)}\big]\nabla^{-2}\partial^i\delta_{mC}^{(1)}}.
\ee
 From \eqref{gs2}--\eqref{15a}, using  \eqref{1}, we find that
\bea
\delta_{g{\rm T}}^{(2)}&=&b_1\delta_{m{\rm C}}^{(2)} +b_2\big[\delta_{m{\rm C}}^{(1)}\big]^2
+2b_1\big[\partial_i\delta_{m{\rm C}}^{(1)}\big]\nabla^{-2}\partial^i\delta_{mC}^{(1)} 
\nonumber\\ \label{16}
&=&b_1\Big[\delta_{m{\rm C}}^{(2)}+2\big[\partial_i\delta_{m{\rm T}}^{(1)}\big] \nabla^{-2}\partial^i\delta_{mC}^{(1)}\Big] +b_2\big[\delta_{m{\rm C}}^{(1)}\big]^2
\eea
which implies
\be \label{17b}
\delta_{g{\rm T}}^{{{({2})}}}=b_1 \delta_{m{\rm T}}^{{{({2})}}} +b_2\big[\delta_{m {\rm T}}^{{{({1})}}}\big]^2.
\ee
 Therefore local-in-mass-density and scale-independent bias in C and T gauge are equivalent up to second order, with the same Eulerian bias coefficients. 
 
We will use the T gauge, since the relation to the Poisson gauge overdensity is simpler for T gauge than C gauge. In Appendix A, we show that
\bea
\delta_{g }^{{{({2})}}}  &=&\delta_{g{\rm T} }^{{{({2})}}}+
(3-b_e) \HH v^{{{({2})}}}  +2(3-b_e)\HH  v^{{{({1})}}} \delta_{g{\rm T}}^{{{({1})}}}  - 2  v^{{{({1})}}} {\delta_{g{\rm T}}^{{{({1})}}\prime}}
 \nonumber\\&&{} 
+\Big[ (b_e-3) \HH' + {b_e' \HH} + (b_e-3)^2 \HH^2 \Big] \big[v^{{{({1})}}}\big]^2 + (b_e-3)\HH   v^{{{({1})}}}  v^{{{({1})}}\prime}
  \nonumber\\&&{} 
 - \left(b_e-3\right) \HH  \nabla^{-2}\bigg[ v^{{{({1})}}} \nabla^2 {v^{{{({1})}}\prime}}
 - {v^{{{({1})}}\prime}}\nabla^2 v^{{{({1})}}} - 6 \partial_i \Phi^{{{({1})}}} \partial^i v^{{{({1})}}} - 6 \Phi^{{{({1})}}} \nabla^2 v^{{{({1})}}}\bigg] .
 \label{dgt2}
\eea
{By \eqref{sb} and \eqref{17b}, this leads to the final expression for the Poisson-gauge galaxy density contrast in the simplest local bias model: 
\bea
\delta_{g }^{{{({2})}}}&=& b_1 \delta_{m{\mathrm{T}}}^{(2)}+ b_2\big[\delta_{m{\mathrm{T}}}^{(1)}\big]^2+\Big[(b_{e}-3)^2\HH^2+{b_{e}'\HH} +(b_{e}-3){\HH'}\Big]\big[v^{(1)}\big]^{2} + (b_{e} - 3)\mathcal{H}v^{(1)}v^{(1)\prime}  +2b_1(3-b_{e})\mathcal{H}v^{(1)}\delta_{m\mathrm{T}}^{(1)}
\nonumber \\&&
 - 2v^{(1)}\Big[b_1{\delta_{m\mathrm{T}}^{(1)\prime}}+b_1'\delta_{m\mathrm{T}}^{(1)}\Big]  + (3-b_{e})\mathcal{H}\nabla^{-2}\bigg[v^{(1)}\nabla^{2}{v^{(1)\prime}}- {v^{(1)\prime}}\,\nabla^{2}v^{(1)} - 6\partial_{i}\Phi^{(1)}\partial^{i}v^{(1)} - 6\Phi^{(1)}\nabla^{2}v^{(1)}\bigg] . \label{pog}
 \eea
 The velocity and metric potential terms ensure gauge-independence on ultra-large scales. Equation \eqref{pog} is the second-order generalisation of \eqref{eq:lineargaugetrans}.}

 \newpage
 \subsection{Observed galaxy number counts in Poisson gauge}\label{sec:linear}

At first order, we replace $\delta_g^{(1)}$ using the bias relations \eqref{sb}--\eqref{1}, and then split  $\Delta_g^{{{({1})}}}$  into Newtonian and GR parts: 
\begin{eqnarray}
 \Delta^{{{({1})}}}_{g{\rm N}} &=& b_1\delta_{m {\rm T}}^{{{({1})}}} - \frac{1}{\HH}  \partial_{\|}^2 v^{{{({1})}}} ,  \label{eq:linearNewtonian}
\\
\Delta^{{{({1})}}}_{g{\rm GR}} &=& \left[ b_e   - 2 \mathcal{Q}  + \frac{2\left( \mathcal{Q}-1 \right)}{\chi \HH} - \frac{\HH'}{\HH^2} \right]\left[ \partial_{\|} v^{{{({1})}}}-  \Phi^{{{({1})}}}\right]+\left(2 \mathcal{Q} -1\right)  \Phi^{{{({1})}}}   + \frac{1}{\HH} \Phi^{{{({1})}}} {'} + \left(3- b_e \right)\HH v^{{{({1})}}}.
\,\label{eq:linearGR}
\end{eqnarray}
The Newtonian part = T-gauge density contrast + {Kaiser} RSD, and the GR part = Doppler + potential + velocity potential. The velocity potential arises from the term in \eqref{eq:lineargaugetrans}, which may be expressed in terms of the metric potential via  \eqref{phi} and \eqref{vm}. The Doppler term in \eqref{eq:linearGR} is the one proportional to the line-of-sight velocity $ \partial_{\|} v^{{{({1})}}}$.

At second order, we use {the gauge-independent bias model \eqref{pog} to replace the Poisson-gauge $\delta_{g }^{{{({2})}}}$
 term in $\Delta_{g }^{{{({2})}}}$.}  The remaining terms in $\Delta_{g }^{{{({2})}}}$ are second-order generalisations of RSD, Doppler and potential terms, together with quadratic couplings amongst all the first-order terms. The quadratic terms encode an interaction between two effects; in Fourier space, they correspond to mode coupling. 
 
{The general equation for $\Delta_g^{(2)}$, including evolution bias and magnification bias, as well as all integrated effects, is given in \cite{Bertacca:2014hwa} {(including recent corrections \cite{Bertacca:2017})}. We include in this general expression our {gauge-independent model of the galaxy bias at second order, \eqref{pog},} and we neglect the terms with integrated contributions. The result  is
\begin{eqnarray} 
\Delta_{g}^{(2)} &=& b_1 \delta_{m{\mathrm{T}}}^{(2)}+ b_2\big[\delta_{m{\mathrm{T}}}^{(1)}\big]^2+\Big[(b_{e}-3)^2\HH^2+{b_{e}'\HH} +(b_{e}-3){\HH'}\Big]\big[v^{(1)}\big]^{2} + (b_{e} - 3)\mathcal{H}v^{(1)}v^{(1)\prime}  +2b_1(3-b_{e})\mathcal{H}v^{(1)}\delta_{m\mathrm{T}}^{(1)}
\nonumber \\&&
 - 2v^{(1)}\Big[b_1{\delta_{m\mathrm{T}}^{(1)\prime}}+b_1'\delta_{m\mathrm{T}}^{(1)}\Big]  + (3-b_{e})\mathcal{H}\nabla^{-2}\bigg[v^{(1)}\nabla^{2}{v^{(1)\prime}}- {v^{(1)\prime}}\,\nabla^{2}v^{(1)} - 6\partial_{i}\Phi^{(1)}\partial^{i}v^{(1)} - 6\Phi^{(1)}\nabla^{2}v^{(1)}\bigg]    \nonumber \\
 &&- \frac{1}{\mathcal{H}}\partial_{\parallel}^{2}v^{(2)}+ (3-b_{e})\mathcal{H}v^{(2)}+ \bigg[b_{e} - 2\mathcal{Q} -\frac{2(1-\mathcal{Q})}{\chi\mathcal{H}}- \frac{\mathcal{H}'}{\mathcal{H}^{2}} \bigg]\left[\partial_{\parallel}v^{(2)}-\Phi^{(2)} \right] +2(\mathcal{Q}-1)\Psi^{(2)}  +\Phi^{(2)} + \frac{1}{\mathcal{H}}{\Psi^{(2)}}' \nonumber \\
&&+ \left[ b_e-2 \Q  -   \frac{\HH'}{\HH^2} - \left(1 - \Q\right) \frac{2}{\chi \HH} \right]\bigg[3  \big[{\Phi}^{(1)}\big]^2  -  \big[ \p_\| v^{(1)} \big]^2+ \p_{\perp i} v^{(1)}  \p^i_{\perp} v^{(1)}  -2  \p_\| v^{(1)}  \Phi^{(1)} 
 \nonumber \\ 
&&- \frac{2}{\HH}\left( \Phi^{(1)}  -  \p_\| v^{(1)}\right) \left(  \Phi^{(1)\prime} - \p_\|^2 v^{(1)}  \right)\bigg] +2\left(2\Q-1 \right)\Phi^{(1)} \delta_g^{(1)} - \frac{2}{\HH} \delta_g^{(1)} \p_\|^2 v^{(1)}  +\frac{2}{\HH}\delta_g^{(1)}  \Phi^{(1)\prime}
\nonumber \\
 && + \left( 4\Q -5 +4 \Q^2 - 4 \frac{\p \Q}{\p \ln \bar L} \right)\big[\Phi^{(1)}\big]^2  
+ \frac{2}{\HH}\left({2\Q}+\frac{\HH'}{\HH^2}\right)\Phi^{(1)} \Phi^{(1)\prime}  
\nonumber\\ 
&& - \frac{2}{\HH}\left({1+ 2\Q}  +\frac{\HH'}{\HH^2}\right)\Phi^{(1)} \p_\|^2 v^{(1)} + \frac{2}{\HH^2}\big[ \Phi^{(1)\prime}  \big]^2 + \frac{2}{\HH^2}\big[\p_\|^2 v^{(1)}  \big]^2  + \frac{2}{\HH^2} \p_\| v^{(1)}  \p_\|^2 \Phi^{(1)}+\frac{4}{ \HH }\p_\| v^{(1)}  \p_\| \Phi^{(1)}
\nonumber\\
&& - \frac{2}{\HH^2} \Phi^{(1)} \p_\|^3 v^{(1)}   -\frac{2}{\HH}\Phi^{(1)} \p_\| \Phi^{(1)}  + \frac{2}{\HH^2}\Phi^{(1)} \frac{\d \Phi^{(1)\prime} }{\d \chi}- \frac{2}{\HH^2} \p_\| v^{(1)} \frac{\d \Phi^{(1)\prime} }{\d \chi}+\frac{2}{\HH} \left(1 
+\frac{\HH'}{\HH^2} \right) \p_\| v^{(1)} \p_\|^2 v^{(1)} 
\nonumber \\
 && - \frac{2}{\HH^2}\Phi^{(1)}  \p_\|^2 \Phi^{(1)}
 +\frac{2}{\HH} \left(1
 -\frac{\HH'}{\HH^2} \right)  \p_\| v^{(1)}  \Phi^{(1)\prime}
   -  \frac{4}{\HH^2}\p_\|^2 v^{(1)}   \Phi^{(1)\prime}  +\frac{2}{\HH}\p_{\perp i} v^{(1)} \p^i_\perp \Phi^{(1)}   -\frac{{4}}{ \HH } \p_{\perp i} v^{(1)}   \p_{\perp}^i \p_\|  v^{(1)}   
\nonumber \\
 &&   +\left(\frac{{4}}{ \chi \HH } -1\right) \p_{\perp i} v^{(1)}    \p_{\perp}^i v^{(1)}   
  +  \frac{2}{\HH^2}\p_\| v^{(1)} \p_\|^3v^{(1)} + \Bigg\{ \bigg[ 4 b_e \Q -2 b_e - 4 \Q   - 8 \Q^2  + 8 \frac{\p \Q}{\p \ln \bar L} +4 \frac{\p \Q}{\p \ln \bar a}   
 \nonumber \\    
&&  + 2  \frac{\HH'}{\HH^2} \left(1 - {2\Q}\right)  + \frac{4}{\chi \HH}\left(\Q -1+2\Q^2- 2\frac{\p \Q}{\p \ln \bar L} \right) \bigg] \Phi^{(1)}  
+2 \bigg[ b_e - 2\Q   -  \frac{\HH'}{\HH^2}  - \frac{2}{\chi \HH} \left(1 - \Q\right)  \bigg] \delta_g^{(1)}
\nonumber  \\
 &&   
- \frac{2}{\HH}  \frac{\d  \delta_g^{(1)} }{\d \chi}  +  \frac{2}{\HH}  \left[ 2 \Q  -   b_e   +  \frac{\HH' }{\HH^2}    +  \frac{2}{\chi \HH}  \left(1 - \Q\right)\right]  \p_\|^2 v^{(1)}    + \frac{2}{\HH}\bigg[ b_e-2    -   \frac{2}{\chi \HH} \left(1 - \Q\right)
\nonumber \\
&&- \frac{\HH' }{\HH^2}\bigg]  \Phi^{(1)\prime}   
  - \frac{4}{\HH} \Q \p_\|  \Phi \Bigg\}\left[\p_\| v^{(1)}- \Phi^{(1)} \right] + \Bigg\{b_e^2-b_e+ \frac{\p b_e}{\p \ln \bar a}+6\Q-4 \Q b_e+4 \Q^2 -4\frac{\p \Q}{\p \ln \bar L}-4\frac{\p \Q}{\p \ln \bar a}   
\nonumber \\
&&+ \frac{6}{\chi} \frac{\HH' }{\HH^3} \left(1 - \Q\right) + \left(1-2 b_e + 4\Q \right) \frac{\HH' }{\HH^2}  -\frac{\HH'' }{\HH^3} +3 \frac{\HH^{\prime 2} }{\HH^4} +  \frac{2}{\chi^2 \HH^2} \bigg(1-\Q +2\Q^2  -2\frac{\p \Q}{\p \ln \bar L}\bigg) + \frac{2}{\chi \HH} \bigg[ 1 - 2b_e - \Q   \nonumber \\
&& + 2b_e \Q  - 4 \Q^2 +4\frac{\p \Q}{\p \ln \bar L} +2 \frac{\p \Q}{\p \ln \bar a} \bigg] \Bigg\} \left[\p_\| v^{(1)}- \Phi^{(1)} \right]^{2}+4\bigg[ \bigg(1-\frac{1}{\chi \HH}\bigg) \p_{\parallel}v^{(1)}- \bigg(2-\frac{1}{\chi \HH}\bigg)\Phi^{(1)}  \bigg]\frac{\p \delta_{g}^{(1)}}{\p \ln{\bar{L}}} .\label{e2}
\end{eqnarray}}

The Newtonian part of \eqref{e2} is formed from the density contrast and {Kaiser} RSD terms and their couplings:\footnote{{Note that the GR correction to $\delta_{m{\rm T}}^{(2)}$ does {\em not} enter the bias term $b_1 \delta_{m{\rm T}}^{(2)}$, as explained in \cite{Dai:2015jaa,dePutter:2015vga,Bartolo:2015qva}.  There is a GR correction to $v^{(2)}$, which we neglect here.}}
\begin{eqnarray} 
\Delta^{{{({2})}}}_{g{\rm N}}
  &=& b_1 \delta_{m{\rm T}}^{{{({2})}}} + {b_2\big[\delta_{m{\rm T}}^{(1)}\big]^2}
  - \frac{1}{\mathcal{H}}\partial_{\parallel}^{2}v^{{{({2})}}}  - 2\frac{b_1}{\mathcal{H}}\bigg[\delta_{mT}^{{{({1})}}}\,\partial_{\parallel}^{2}v^{{{({1})}}} + \partial_{\parallel}v^{{{({1})}}}\,\partial_{\parallel}\delta_{mT}^{{{({1})}}}\bigg]
  + \frac{2}{\mathcal{H}^{2}}\bigg[\big[\partial_{\parallel}^{2}v^{{{({1})}}}\big]^{2} + \partial_{\parallel}v^{{{({1})}}}\,\partial_{\parallel}^{3}v^{{{({1})}}}\bigg]. \label{eq:SecondorderNewtonian}
 \eea
The remaining terms form the GR correction:
 \bea
\Delta_{g\mathrm{GR}}^{{{({2})}}} &=& \HH(3-b_{e})v^{(2)} + \Big[(9-6b_{e}+b_{e}^{2})\HH^2+{b_{e}'\HH} +(b_{e}-3){\HH'}\Big]\big[v^{(1)}\big]^{2} + (b_{e} - 3)\mathcal{H}v^{(1)}{v^{(1)\prime}}  
 \nonumber \\
&& - (b_{e} - 3)\mathcal{H}\nabla^{-2}\bigg[v^{(1)}\nabla^{2}{{v^{(1)\prime}}}- {{v^{(1)\prime}}}\,\nabla^{2}v^{(1)} - 6\partial_{i}\Phi^{(1)}\partial^{i}v^{(1)} - 6\Phi^{(1)}\nabla^{2}v^{(1)}\bigg] +2(3-b_{e})b_{1}\mathcal{H}v^{(1)}\delta_{m{\rm T}}^{(1)}  \nonumber\\
&& - 2v^{(1)}\left(b_{1}'\delta_{m{\rm T}}^{(1)} + b_{1}{\delta_{m{\rm T}}^{(1)\prime}}\right) + \bigg[b_{e}-2Q-\frac{2(1-\Q)}{\chi\HH}-\frac{\HH'}{\HH^{2}}\bigg]\p_{\parallel}v^{(2)} + \bigg[1-b_{e}+2\Q+\frac{2(1-\Q)}{\chi\HH}+\frac{\HH'}{\HH^{2}}\bigg]\Phi^{(2)} \nonumber \\
&&  - 2(1-\mathcal{Q})\Psi^{(2)} + \frac{1}{\mathcal{H}}{\Psi^{(2)\prime}} + \frac{2}{\HH}\bigg[ b_{1}{\delta_{m{\rm T}}^{(1)\prime}}\,\p_{\parallel}v^{(1)}+(f-2+2\Q)\Phi^{(1)}\p_{\parallel}\Phi^{(1)} +(2-f-2\Q)\p_{\parallel}v^{(1)}\p_{\parallel}\Phi^{(1)}  \nonumber \\
&&  - b_{1}\Phi^{(1)} {\delta_{m{\rm T}}^{(1)\prime}} + b_{1}\Phi^{(1)}\p_{\parallel}\delta_{m{\rm T}}^{(1)} - 2\p_{i}v^{(1)}\p_{\parallel}\p^{i}v^{(1)} + \p_{i}v^{(1)}\p^{i}\Phi^{(1)}\bigg] + \frac{2}{\HH^{2}}\bigg[\p_{\parallel}v^{(1)}\p_{\parallel}^{2}\Phi^{(1)} - \Phi^{(1)}\p_{\parallel}^{2}\Phi^{(1)} - \Phi^{(1)}\p_{\parallel}^{3}v^{(1)}\bigg]  \nonumber \\
&&   - 2(3-b_{e})v^{(1)}\p_{\parallel}^{2}v^{(1)} + 2\bigg[b_{1}\bigg(b_{e}-2\Q-\frac{2(1-\Q)}{\chi\HH} - \frac{\HH'}{\HH^{2}} \bigg)+ \frac{b_{1}'}{\HH}+ 2\bigg(1-\frac{1}{\chi\HH}\bigg)\frac{\p b_{1}}{\p \ln{\bar{L}}}\bigg]\delta_{m{\rm T}}^{(1)}\p_{\parallel}v^{(1)}  \nonumber \\
&&  +\frac{2}{\HH}\bigg[3-2b_{e}+{4}\Q+\frac{4(1-\Q)}{\HH\chi}+\frac{3\HH'}{\HH}\bigg]\p_{\parallel}v^{(1)}\p_{\parallel}^{2}v^{(1)}  + 2\bigg[b_{1}\bigg(f-2-b_{e}+4\Q+\frac{2(1-\Q)}{\chi\HH}+\frac{\HH'}{\HH^{2}}\bigg) -\frac{{b_{1}'}}{\HH} \nonumber \\
&&-2\bigg(2-\frac{1}{\chi\HH}\bigg)\frac{\p b_{1}}{\p \ln{\bar{L}}}\bigg]\Phi^{(1)}\delta_{m{\rm T}}^{(1)} + \bigg[b_{e}-1-2\Q-\frac{2(1-\Q)}{\chi\HH}-\frac{\HH'}{\HH^{2}}\bigg]\p_{i}v^{(1)}\p^{i}v^{(1)} + \frac{2}{\HH}\bigg[{1}-2f+2b_{e}-{6}\Q \nonumber \\
&& -\frac{4(1-\Q)}{\chi\HH}-\frac{3\HH'}{\HH^{2}}\bigg]\Phi^{(1)}\p_{\parallel}^{2}v^{(1)} + \mathcal{A}\big[\Phi^{(1)}\big]^{2} + \mathcal{B}v^{(1)}\p_{\parallel}v^{(1)} + \mathcal{C}\Phi^{(1)} v^{(1)} + \mathcal{D}\Phi^{(1)}\p_{\parallel}v^{(1)} + \mathcal{E}\big[\p_{\parallel}v^{(1)}\big]^{2} .
 \label{eq:secondorderGRA}
\end{eqnarray}
The background coefficients in the last line are
\begin{eqnarray}
\mathcal{A} &=& {-3} +2f\bigg({2}-2b_{e}+{4}\Q+\frac{4(1-\Q)}{\chi\HH} +\frac{2\HH'}{\HH^{2}}\bigg) -\frac{2f'}{\HH} +b_{e}^{2}+ 6b_{e}-8b_{e}\Q+4\Q+16\Q^{2} -16\frac{\p \Q}{\p \ln\bar{L}} \nonumber\\ 
&& -8\frac{\Q'}{\HH} + \frac{b_{e}'}{\HH}+\frac{2}{\chi^{2}\HH^{2}}\bigg(1-\Q+2\Q^{2}-2\frac{\p \Q}{\p \ln{\bar{L}}}\bigg)- \frac{2}{\chi\HH}\bigg[4+2b_{e}-2b_{e}\Q-4\Q+8\Q^{2}-\frac{3\HH'}{\HH^{2}}(1-\Q) \nonumber \\
&&{} - 8\frac{\p \Q}{\p \ln{\bar{L}}} - 2\frac{\Q'}{\HH}\bigg] + \frac{\HH'}{\HH^{2}}\bigg(-8-2b_{e}+{8}\Q+\frac{3\HH'}{\HH^{2}}\bigg) - \frac{\HH''}{\HH^{3}}, \label{bca}\\
\mathcal{B} &=& 2\HH\bigg[-3+4b_{e}+2b_e\frac{(1-\Q)}{\chi\HH}-b_{e}^{2}+2b_{e}\Q -6\Q-\frac{b_{e}'}{\HH}-\frac{6(1-\Q)}{\chi\HH}+2\bigg(1-\frac{1}{\chi\HH}\bigg)\frac{\Q'}{\HH}\bigg], \\
\mathcal{C} &=& 2\HH\bigg[-3+f(3-b_{e})-3b_{e}-2b_e\frac{(1-\Q)}{\chi\HH}+\frac{b_{e}'}{\HH}+b_{e}^{2}-4b_{e}\Q+12\Q+\frac{6(1-\Q)}{\chi\HH}-2\bigg(2-\frac{1}{\chi\HH}\bigg)\frac{\Q'}{\HH}\bigg], \\
\mathcal{D} &=& 4 +2f\bigg[-3+f+2b_{e}-3\Q-\frac{4(1-\Q)}{\chi\HH}-\frac{2\HH'}{\HH^{2}}\bigg] +\frac{2f'}{\HH}-6b_{e}-2b_{e}^{2}+12b_{e}\Q-{8}\Q-16\Q^{2}+16\frac{\p \Q}{\p \ln{\bar{L}}}  \nonumber \\
&&+12\frac{\Q'}{\HH} -2\frac{b_{e}'}{\HH} -\frac{4}{\chi^{2}\HH^{2}}\bigg(1-\Q+2\Q^{2}-2\frac{\p \Q}{\p \ln{\bar{L}}}\bigg) - \frac{4}{\chi \HH}\bigg[-1 -2b_{e}+2b_{e}\Q+\Q-6\Q^{2}+\frac{3\HH'}{\HH^{2}}(1-\Q)  \nonumber \\
&& +6\frac{\p \Q}{\p \ln{\bar{L}}} + 2\frac{\Q'}{\HH}\bigg]+ \frac{2\HH'}{\HH^{2}}\bigg(3+2b_{e}-6\Q-\frac{3\HH'}{\HH^{2}}\bigg) + \frac{2\HH''}{\HH^{3}} , \\
\mathcal{E} &=&  -4-b_{e}+b_{e}^{2}-4b_{e}\Q+ {6}\Q+4\Q^{2}-4\frac{\p \Q}{\p \ln{\bar{L}}} - 4\frac{\Q'}{\HH} + \frac{b_{e}'}{\HH} + \frac{2}{\chi^{2}\HH^{2}}\bigg(1-\Q+2\Q^{2}-2\frac{\p \Q}{\p \ln{\bar{L}}}\bigg) \nonumber \\
&& + \frac{2}{\chi\HH}\bigg[3-2b_{e}+2b_{e}\Q-3\Q-4\Q^{2}+\frac{3\HH'}{\HH^{2}}(1-\Q) + 4\frac{\p \Q}{\p \ln{\bar{L}}} + 2\frac{\Q'}{\HH}\bigg] + \frac{\HH'}{\HH^{2}}\bigg(3-2b_{e}+{4}\Q+\frac{3\HH'}{\HH^{2}}\bigg) - \frac{\HH''}{\HH^{3}}.\nonumber \\
&& \label{bce}
\end{eqnarray}

{In deriving \eqref{eq:SecondorderNewtonian}--\eqref{bce} from \eqref{e2}, we used the following:\\ \\
(a) eliminate $\d/\d\chi$ using \eqref{chi}, and $\partial_{\perp i}$ using \eqref{rtder};\\
(b) show, {using the commutator relation} $\big[ \partial_{\perp i},\partial_{\|}\big]=\chi^{-1}\partial_{\perp i}$, that 
\be
\partial_{\perp i}v^{(1)}\,\partial_{\perp}^i\partial_{\|}v^{(1)}=\partial_iv^{(1)}\,\partial_{\|}\partial^iv^{(1)}
-\partial_{\|}v^{(1)}\, \partial_{\|}^2v^{(1)}+{1\over\chi}\Big[ \partial_iv^{(1)}\,\partial^iv^{(1)}-\big[\partial_{\|}v^{(1)} \big]^2\Big];
\ee 
(c)  express $\delta_{g}^{{{({1})}}}$ in terms of $\delta_{m{\rm T}}^{{{({1})}}}$ and $v^{(1)}$, using \eqref{eq:lineargaugetrans} and \eqref{1};\\
(d) rewrite
the term  from the perturbation of the magnification bias, using {\eqref{sb}--\eqref{1},} as
\be
{\partial\delta_g^{{{({1})}}} \over \partial \ln \bar L}   
={\partial b_1\over\partial \ln \bar L} \,{\delta_{m{\rm T}}^{{{({1})}}}} -{\partial b_e\over\partial \ln \bar L}\,\HH v^{{{({1})}}}
={\partial b_1\over\partial \ln \bar L} \,{\delta_{m{\rm T}}^{{{({1})}}}}  + \Q' v^{{{({1})}}},
\label{dq}
\ee

where the second equality uses \eqref{Q}, \eqref{be} {and $\partial/\partial\ln a=\HH^{-1}\partial/\partial\eta$}.}\\

{In summary: we have used the general formula for $\Delta_{g}^{(2)}$ in Poisson gauge, given in  \cite{Bertacca:2014hwa}, neglecting the terms with line-of-sight integrals, to derive \eqref{eq:SecondorderNewtonian}--\eqref{bce}. In these equations we have broken down the highly complex formula in  \cite{Bertacca:2014hwa} into simple parts, facilitating analytical and then numerical analysis.
 Our new contribution is to determine the Poisson-gauge $\delta_g^{(2)}$ via a simple local-in-mass-density model of bias \eqref{pog}, that is gauge independent and valid on ultra-large scales.\footnote{{Three groups have computed $\Delta_{g}^{(2)}$ -- in \cite{Bertacca:2014hwa,Bertacca:2014dra,Bertacca:2014wga}, \cite{Yoo:2014sfa} and \cite{DiDio:2014lka}. All have used different formalisms. The collective task of cross-checking these independent results has been initiated but is not complete, even in the simplest case with no integrated contributions and $b_e=0={\cal Q}$.}}}

\newpage
\section{Galaxy number overdensity in Fourier space and the bispectrum}\label{sec:generaldef}

We will only consider correlations at the same {observed} redshift. At fixed redshift $z$, the perturbative variables depend on $\bm n$ and can be computed  in Fourier space at fixed $\eta(z)$. {With $\bm n$ and $z$ fixed, we transform $\bm x =[\eta_0-\eta(z)]\bm n +\bm x_0 \to \k$, which is equivalent to transforming over all observer positions $\bm x_0$.}
Our Fourier  convention is 
\begin{equation}\label{eq:STF1}
f({\bm {x}}) = \int \frac{\d^{3} k}{(2 \pi)^{3}}\, {\rm e}^{{\rm i} {\bm{k} \cdot \bm{x}}}f({\bm{k}}) , \quad
f({\bm{k}})=\int \d^{3} x\, {\rm e}^{-{\rm i} {\bm{k} \cdot \bm{x}}}f({\bm{x}})=\int\frac{\d^3k'}{(2\pi)^3}(2\pi)^3\delta^{D}({\bm{k}-\bm{k}'})f(\bm{k}') ,
\end{equation}
where we suppress the redshift dependence. The transform of a product $h({\bm{x}}) = g({\bm{x}}) f({\bm{x}})$ leads to a convolution  in Fourier space
\begin{eqnarray}
h({\bm{k}})&=&\int\frac{\d^3 k_1}{(2\pi)^3}\frac{\d^3k_2}{(2\pi)^3}f({\bm{k}}_1) g(\bm{k}_{2})(2\pi)^3 \delta^{D}\left( {\bm{k}_{1}+\bm{k}_{2}-\bm{k}}\right).
\end{eqnarray}

For notational convenience we write the T-gauge matter density contrast as
\be
{\delta_{m{\rm T}}\equiv\delta}=\delta^{(1)}+{1\over2}\delta^{(2)},
 \ee
from now on.

{At second order, the matter density contrast and the velocity and metric potentials are given in a Newtonian approximation by \cite{Bernardeau:2001qr}:}
\begin{eqnarray}
\delta^{{{({2})}}}(\bm{k}) &=& \int \frac{\d^3 k_1}{(2\pi)^3}\frac{\d^3 k_2}{(2\pi)^3}\delta^{{{({1})}}}(\bm{k}_{1}) \delta^{{{({1})}}}(\bm{k}_{2})  {F_2}(\bm{k}_{1},\bm{k}_{2})
(2\pi)^3\delta^D\left(\bm{k}_{1} + \bm{k}_{2} - \bm{k}\right),  \label{nd2} \\
 v^{{{({2})}}}(\bm{k}) &=& f\frac{\HH}{k^2} \int \frac{\d^3 k_1}{(2\pi)^3}\frac{\d^3 k_2}{(2\pi)^3}\delta^{{{({1})}}}(\bm{k}_{1}) \delta^{{{({1})}}}(\bm{k}_{2})  G_2(\bm{k}_{1},\bm{k}_{2})
(2\pi)^3\delta^D\left(\bm{k}_{1} + \bm{k}_{2} - \bm{k}\right), \label{nv2}
\\
\Phi^{{{({2})}}}(\k) &=& \Psi^{{{({2})}}}(\bm{k}) =-{3\over2}\Omega_m\frac{\mathcal{H}^2 }{k^2} \delta^{{{({2})}}}(\bm{k}) .
\label{np2}
\end{eqnarray}
The kernels for the dark matter and peculiar velocity perturbations in a matter-dominated model are 
\begin{eqnarray} 
F_{2}(\bm{k}_{1}, \bm{k}_{2}) &=& \frac{10}{7} + \frac{\bm{k}_{1} \cdot \bm{k}_{2}}{k_{1}k_{2}}\bigg(\frac{k_{1}}{k_{2}} + \frac{k_{2}}{k_{1}}\bigg) + \frac{4}{7}\bigg(\frac{\bm{k}_{1} \cdot \bm{k}_{2}}{k_{1}k_{2}}\bigg)^{2}  ,
\label{f2} \\
G_{2}(\bm{k}_{1}, \bm{k}_{2}) &=& \frac{6}{7} + \frac{\bm{k}_{1} \cdot \bm{k}_{2}}{k_{1}k_{2}}\bigg(\frac{k_{1}}{k_{2}} + \frac{k_{2}}{k_{1}}\bigg) + \frac{8}{7}\bigg(\frac{\bm{k}_{1} \cdot \bm{k}_{2}}{k_{1}k_{2}}\bigg)^{2}. \label{g2}
\end{eqnarray}
{The corrections to these kernels from the presence of $\Lambda$ are small  \cite{Tram:2016cpy}, and we neglect them. Within the same approximation, we have $\delta^{(2)}\propto D^2 \delta^{(2)}_0$, so that $\delta^{(2)\prime}=2f\HH \delta^{(2)}$.
Then it follows from \eqref{np2} that
\be
\Phi^{(2)\prime}=(2f-1)\HH\,\Phi^{(2)}. \label{pp2}
\ee }

We write  $\Delta_g^{(1,2)}$ in terms of kernels:
\begin{eqnarray} \label{eq:densityinfourierspace1}
\Delta_{g}^{{{({1})}}}(\bm{k}_{2}) &=& \int \frac{\d^{3}k_1}{(2\pi)^{3}}\mathcal{K}^{{{({1})}}}(\bm{k}_{1})\delta^{{{({1})}}}(\bm{k}_{1})(2\pi)^3\delta^{D}(\bm{k}_{1}-\bm{k}_{2}) , \\
\Delta_{g}^{{{({2})}}}(\bm{k}_{3}) &=& \int \frac{\d^{3}k_{1}}{(2\pi)^{3}}\frac{\d^{3}k_{2}}{(2\pi)^{3}}\,\mathcal{K}^{{{({2})}}}(\bm{k}_{1}, \bm{k}_{2}, \bm{k}_{3})\delta^{{{({1})}}}(\bm{k}_{1})\delta^{{{({1})}}}( \bm{k}_{2})(2\pi)^3\delta^{D}(\bm{k}_{1} + \bm{k}_{2} - \bm{k}_{3})  - \delta^{(D)}({\k}_3) \big\<\Delta_g^{{{({2})}}}\big\>,
\label{eq:densityinfourierspace2}
\end{eqnarray}
and we split the kernels into Newtonian and GR parts, $\mathcal{K}^{(1,2)}=\mathcal{K}^{(1,2)}_{\rm N}+\mathcal{K}^{(1,2)}_{\rm GR}$. In \eqref{eq:densityinfourierspace2}, we subtracted off the ensemble average of $\Delta_g$:
\begin{eqnarray}\label{eq;onepointfunc}
\big\<\Delta_g^{{{({2})}}}\big\> =  \int {{\d^3 k_1 \over (2\pi)^3}\,P(k_1)\mathcal{K}^{{{({2})}}}(\bm{k}_{1}, -\bm{k}_{1}, 0) ,}
\end{eqnarray}
in order to ensure that $\<\Delta_g\> = 0$. Here $P(k)\equiv P_{\delta^{(1)}}(k)$ is the linear matter power spectrum.

By  \eqref{eq:linearNewtonian} and \eqref{eq:linearGR}, the linear order kernel is given by
\begin{eqnarray}
\mathcal{K}^{{{({1})}}}_{\rm{N}}({\k})& =&   b_{1} + f\mu^{2}\,,\qquad  {
\mathcal{K}^{{{({1})}}}_{\rm{GR}}({\k}) ={\rm i} {\mu\over k} \gamma_1+\frac{\gamma_2}{k^{{2}}} , } \qquad \mu =  \hat{\bm{k}} \cdot {\bm{n}}, \label{k1}
\end{eqnarray}	
where $\gamma_1$ and $\gamma_2$ are redshift dependent:
\begin{eqnarray}
{\gamma_1\over \mathcal{H}} &=&  { f}\bigg[b_{e}  - 2\mathcal{Q} - \frac{2(1-\mathcal{Q})}{\chi\mathcal{H}}- \frac{\mathcal{H}'}{\mathcal{H}^{2}} \bigg] , \label{ga1}
\\
\frac{\gamma_{2}}{\mathcal{H}^{2}} &=& f(3-b_{e}) +\frac{3}{2}\Omega_{m} \left[2+b_{e}-f-4\mathcal{Q}-2 \frac{\left(1- \mathcal{Q} \right)}{\chi\mathcal{H}}-\frac{\mathcal{H}'}{\mathcal{H}^2}\right]. \label{ga2}
\end{eqnarray}

At second order, the Newtonian part of the kernel is 
\begin{eqnarray}\label{eq:FourierNewtonian}
{\mathcal{K}^{{{({2})}}}_{\rm{N} }(\bm{k}_{1}, \bm{k}_{2},{\k_3})} &=& b_{1}F_{2}(\bm{k}_{1}, \bm{k}_{2}) + b_{2} + fG_{2}(\bm{k}_{1}, \bm{k}_{2})\mu_{3}^{2}
\nonumber\\ &&{}
+  f^2{\mu_1\mu_2 \over k_1k_2}\big( \mu_1k_1+\mu_2k_2\big)^2
+
b_1{f\over k_1k_2}\Big[ \big(\mu_1^2+\mu_2^2 \big)k_1k_2+\mu_1\mu_2\big(k_1^2+k_2^2 \big) \Big],
\end{eqnarray}
where  $\mu_{i} =  \hat{\bm{k}}_i \cdot {\bm{n}}$. The second line in \eqref{eq:FourierNewtonian} is the nonlinear {Kaiser} RSD contribution \cite{Verde:1998zr,Scoccimarro:1999ed}.

The GR part follows from \eqref{eq:secondorderGRA}, {after transformation to Fourier space. The details, with all the necessary transforms, are given in Appendix B, and they lead to the GR kernel:}
\begin{eqnarray}\label{e8}
\mathcal{K}^{{{({2})}}}_{\mathrm{GR}}(\bm{k}_{1}, \bm{k}_{2}, \bm{k}_{3}) &=& \frac{1}{k_{1}^{2}k_{2}^{2}}\bigg\{\Gamma_{1} + {\rm i}\left(\mu_{1}k_{1} + \mu_{2}k_{2}\right)\Gamma_{2} + \frac{k_{1}^2k_{2}^2}{k_{3}^2} \Big[F_{2}(\bm{k}_{1}, \bm{k}_{2})\,\Gamma_{3} + G_{2}(\bm{k}_{1}, \bm{k}_{2})\,\Gamma_{4} \Big] \nonumber \\&&{}
 + \left(\mu_{1}\mu_{2}k_{1}k_{2}\right)\Gamma_{5} + \left(\bm{k}_{1}\cdot \bm{k}_{2}\right)\Gamma_{6} + \left(k_{1}^{2} + k_{2}^{2}\right)\Gamma_{7} + \left(\mu_{1}^{2}k_{1}^{2} + \mu_{2}^{2}k_{2}^{2}\right)\Gamma_{8} 
 \nonumber \\&&{}
 +  {\rm i}\bigg[\left(\mu_{1}k_{1}^{3} + \mu_{2}k_{2}^{3}\right)\Gamma_{9}  
+ \left(\mu_{1}k_{1} + \mu_{2}k_{2}\right)\left(\bm{k}_{1} \cdot \bm{k}_{2}\right)\Gamma_{10}  + k_{1}k_{2}\left(\mu_{1}k_{2} + \mu_{2}k_{1}\right)\Gamma_{11}   
\nonumber \\&&{}
+ \left(\mu_{1}^{3}k_{1}^{3}+ \mu_{2}^{3}k_{2}^{3}\right)\Gamma_{12}
 + \mu_{1}\mu_{2}k_{1}k_{2}\left(\mu_{1}k_{1} + \mu_{2}k_{2}\right)\Gamma_{13} + \mu_{3}\frac{k_{1}^{2}k_{2}^{2}}{k_{3}}\,G_{2}(\bm{k}_{1}, \bm{k}_{2})\,\Gamma_{14}\bigg] \bigg\}\,,
\end{eqnarray}
where  the $\Gamma_I(z)$ are  given in Appendix {C}.  

{We have ordered the $\Gamma_I$ according to the powers of $\HH/k$, starting with the ${\cal O}(\HH^4/k^4)$ term and ending with the ${\cal O}(\HH/k)$ terms.}
This is our key result -- transforming the highly complicated second-order GR projection corrections given by  \eqref{eq:secondorderGRA} into a manageable Fourier-space kernel \eqref{e8}. 
In the special case $b_e=0={\cal Q}$,  \eqref{eq:secondorderGRA} reduces to the form given  in \cite{Umeh:2016nuh}. When $b_e,{\cal Q}$ are nonzero, the $\Gamma_I$ become much more complicated.

In Fourier space, the observed galaxy bispectrum $B_g$ at fixed redshift is given by
\begin{equation} 
\big\langle \Delta_{g}( \bm{k}_{1}) \Delta_{g}( \bm{k}_{2}) \Delta_{g}( \bm{k}_{3})\big \rangle = (2\pi)^{3}B_{g}(  \bm{k}_{1},  \bm{k}_{2},  \bm{k}_{3})\delta^{D}( \bm{k}_{1}+ \bm{k}_{2}+ \bm{k}_{3}).
\end{equation}
 At second order, the only combinations of terms that contribute at tree-level  are
\begin{eqnarray} 
2\big \langle \Delta_{g}( \bm{k}_{1}) \Delta_{g}( \bm{k}_{2}) \Delta_{g}( \bm{k}_{3})\big \rangle &=&\big\langle \Delta_{g}^{{{({1})}}}(\bm{k}_{1}) \Delta_{g}^{{{({1})}}}(\bm{k}_{2}) \Delta_{g}^{{{({2})}}}( \bm{k}_{3})\big \rangle  + \text{2\;cyc.\;perm.}
\\
 &=& \big\<\Delta^{{{({1})}}}_{g\rm{N}}( \bm{k}_{1})\Delta^{{{({1})}}}_{g\rm{N}}( \bm{k}_{2})\Delta^{{{({2})}}}_{g\rm{N}}( \bm{k}_{3})\big\>
+\big \<\Delta^{{{({1})}}}_{g\rm{GR}}( \bm{k}_{1})\Delta^{{{({1})}}}_{g\rm{GR}}( \bm{k}_{2})\Delta^{{{({2})}}}_{g\rm{GR}}( \bm{k}_{3})\big\> 
\nonumber\\ &&{}
+ \big\<\Delta^{{{({1})}}}_{g\rm{N}}( \bm{k}_{1})\Delta^{{{({1})}}}_{g\rm{N}}( \bm{k}_{2})\Delta^{{{({2})}}}_{g\rm{GR}}( \bm{k}_{3})\big\> 
+ \big\<\Delta^{{{({1})}}}_{g\rm{GR}}( \bm{k}_{1})\Delta^{{{({1})}}}_{g\rm{GR}}( \bm{k}_{2})\Delta^{{{({2})}}}_{g\rm{N}}( \bm{k}_{3})\big\>
\nonumber\\ &&{}
+ 2\left[\big\<\Delta^{{{({1})}}}_{g\rm{N}}( \bm{k}_{1})\Delta^{{{({1})}}}_{g\rm{GR}}( \bm{k}_{2})\Delta^{{{({2})}}}_{g\rm{N}}( \bm{k}_{3})\big\> + \big\<\Delta^{{{({1})}}}_{g\rm{N}}( \bm{k}_{1})\Delta^{{{({1})}}}_{g\rm{GR}}( \bm{k}_{2})\Delta^{{{({2})}}}_{g\rm{GR}}( \bm{k}_{3})\big\>\right]
\nonumber\\  &&{}
+ \text{2\;cyc.\;perm.},
\label{eq:defbispec}
\end{eqnarray}
where the factors of $2$ arise from the factor $1/2$ in the perturbative expansion of $\Delta_g$. In the second equality, we have further separated the bispectrum  into purely Newtonian and purely GR parts (first line), and  cross-correlations between Newtonian and GR terms (following lines). The cross-correlation terms {become important on smaller scales than the pure GR term}.

The full expression for the galaxy bispectrum in terms of kernels follows from \eqref{eq:defbispec} as:
\begin{eqnarray}\label{eq:bispectrumGen}
B_{g}( \bm{k}_{1},  \bm{k}_{2},  \bm{k}_{3}) &=&
 \bigg[\mathcal{K}^{{{({1})}}}_{\rm{N}}(\bm{k}_{1}) \mathcal{K}^{{{({1})}}}_{\rm{N}}(\bm{k}_{2}) \mathcal{K}^{{{({2})}}}_{\rm{N}}(\bm{k}_{1},  \bm{k}_{2},\bm{k}_{3}) +
 \mathcal{K}^{{{({1})}}}_{\rm{GR}}(\bm{k}_{1}) \mathcal{K}^{{{({1})}}}_{\rm{GR}}(\bm{k}_{2}) \mathcal{K}^{{{({2})}}}_{\rm{GR}}(\bm{k}_{1},  \bm{k}_{2},\bm{k}_{3})  
 \nonumber\\ &&{}
+\mathcal{K}^{{{({1})}}}_{\rm{N}}(\bm{k}_{1}) \mathcal{K}^{{{({1})}}}_{\rm{N}}(\bm{k}_{2}) \mathcal{K}^{{{({2})}}}_{\rm{GR}}(\bm{k}_{1},  \bm{k}_{2},\bm{k}_{3})  
 + \mathcal{K}^{{{({1})}}}_{\rm{GR}}(k_1)\mathcal{K}_{\rm{GR}}^{{{({1})}}}(k_2) \mathcal{K}^{{{({2})}}}_{\rm{N}}(\bm{k}_{1},  \bm{k}_{2},\bm{k}_{3}) 
\nonumber \\ &&{}
 +2\mathcal{K}^{{{({1})}}}_{\rm{N}}(\bm{k}_{1})\mathcal{K}^{{{({1})}}}_{\rm{GR}}(\bm{k}_{2})\left\{\mathcal{K}^{{{({2})}}}_{\rm{N}}(\bm{k}_{1},  \bm{k}_{2},\bm{k}_{3}) 
 + \mathcal{K}^{{{({2})}}}_{\rm{GR}}(\bm{k}_{1},  \bm{k}_{2},\bm{k}_{3})\right\}
\bigg]P(k_{1})P(k_{2})  +  \text{2 cyc. perm.}
\end{eqnarray}
The bispectrum in the Newtonian approximation is
\begin{equation} \label{eq:Newtonianbispecttrum}
B_{g{\rm N}}( \bm{k}_{1},  \bm{k}_{2},  \bm{k}_{3}) = \mathcal{K}^{{{({1})}}}_{\rm{N}}(\bm{k}_{1}) \mathcal{K}^{{{({1})}}}_{\rm{N}}(\bm{k}_{2}) \mathcal{K}^{{{({2})}}}_{\rm{N}}(\bm{k}_{1},  \bm{k}_{2},\bm{k}_{3}) P(k_{1})P(k_{2}) +  \text{2 cyc. perm.}
\end{equation}
 All other terms in \eqref{eq:bispectrumGen} are GR corrections, i.e., they vanish if the GR projection effects are neglected.

Calculation of the galaxy bispectrum including all the GR terms leads to a  complex-valued function.  We split  \eqref{eq:bispectrumGen} into real and imaginary parts  $B_{g} = B_{g}^{\rm{R}} + {\rm i}\, B_{g}^{\rm{I}}$ and compute the  absolute value of the galaxy bispectrum, {given by}  $|B_{g}|^2 = (B_{g}^{\rm R})^2 + ({B_{g}^{\rm{I}}})^2$.

There are four different angles implicit in (\ref{eq:bispectrumGen}):
\begin{itemize} 
\item[]   three {$\theta_i$} between the observer line of sight and the mode vectors (with cosines {$\mu_i=\cos\theta_i=\hat\k_i\cdot\n$})
\item[] +  one of the angles {$\theta_{ij}$ between $\k_i$ and $\k_j$}  (with cosines
{${\mu_{ij} = \cos\theta_{ij}=\hat\k_{i} \cdot \hat\k_{j}}$}). 
\end{itemize}

Two of the $\mu_i$ are independent, since 
$\mu_{1}k_{1} + \mu_{2}k_{2}+\mu_{3}k_{3}=0 $, where $k_3 = |\bm{k}_{1} + \bm{k}_{2}|$. 
Two of the $\mu_{ij}$ can be determined by the third  via trigonometric identities.
Finally, one of the two remaining $\mu_i$ may be expressed in terms of the other one and the choice of independent $\mu_{ij}$,  using the trigonometric addition formula. If we choose $\mu_1$ and $\mu_{12}$, then
\begin{equation}
{\mu_{2} ={\mu_{1}\mu_{12}} \pm \sqrt{1 - \mu_{1}^{2} }\sqrt{1 - \mu_{12}^{2}}\cos \phi}\,,
\end{equation}
where {$\mu_{12}$  can be determined from the $k_i$. Here $\phi$ is the azimuthal angle, characterizing the orientation of the triangle in Fourier space, and the $\pm$ arises due to invariance under reflection of $\n$ about $\hat\k_2$ in their plane.} 

Implementing these conditions, the  galaxy bispectrum is a function of $\mu_{1}$ and $\phi$, together with the magnitudes of the three mode vectors.
The dependence of $B_g$ on $\mu_{1}$ and $\phi$ may  be  expanded in spherical harmonics: 
\bea
B_g({k}_1,{k}_2,{k}_3,\mu_{1},\phi) &=& \sum_{\ell = 0}
\sum_{m = -\ell}^{\ell}B_g^{\ell m}({k}_1,{k}_2,{k}_3) Y_{\ell m }(\mu_1, \phi), 
\eea
where the multipoles of $B_g$ are given by
\bea
\label{eq:bispectrummultipoles}
B_{g}^{\ell m}({k}_1,{k}_2,{k}_3) &=& \frac{(2\ell +1)}{4\pi} \int_0^{2\pi} \d\phi\int_{-1}^{1}\d\mu_{1}\, B_g({k}_1,{k}_2,{k}_3,\mu_{1}, \phi) Y_{\ell m }^{*}(\mu_1, \phi).
\eea
This can be compared to the Legendre multipole expansion of the galaxy power spectrum
\begin{equation}\label{eq:harmonicexp}
P_{g}(k,\mu) = \sum_{\ell = 0}^{\ell_{\text{max}}}P_{g}^{\ell}(k) \mathcal{L}_{\ell}(\mu)
 \quad \text{with}\quad
P_{g}^{\ell}(k) = \frac{(2\ell+1)}{2} \int_{-1}^{1}\d\mu\, P_{g}(k,\mu) \mathcal{L}_{\ell}(\mu)\,.
\end{equation}
Note that we can also expand the bispectrum in Associated Legendre polynomials and still recover the multipoles as given in  \eqref{eq:bispectrummultipoles}.  

Typically, only the $m=0$ multipoles of $B_g$ are considered, and we will do this, so that $B_g=B_g({k}_1,{k}_2,{k}_3,\mu_{1})$. In fact, this does not lose much information \cite{Gagrani:2016rfy}. 
For the monopole, we use the shorthand $B^0_g \equiv B^{00}_g$.

\newpage
\section{Numerical Results }\label{sec4}

In order to illustrate quantitatively the imprint of GR effects on the galaxy bispectrum, we specialise to an isosceles configuration, with
\be
k_1=k_2\equiv k,~~~ {k_3= k\sqrt{2(1+\mu_{12})}}.
\ee
We evaluate the following cases:{
\bea
\mbox{radial:}~ \mu_{1} = 1\to B^\parallel_g,\qquad  \mbox{transverse:}~\mu_{1} =0\to B^\perp_g, \qquad \mbox{monopole:}~\int \d\mu_1\to B^0_g.
\eea  }
For redshifts and astrophysical parameters, we choose:
\be
z =1.0,\,1.5, \qquad b_1(z)= \sqrt{1+z},~~ {b_2(z)=   -0.1 \sqrt{1+z}}, \qquad b_e=0={\cal Q},
\ee
where the galaxy bias parameters {are similar to}  \cite{Pollack:2013alj}.

In each case, we compare the Newtonian prediction \eqref{eq:Newtonianbispecttrum} for the galaxy bispectrum,  to the GR prediction \eqref{eq:bispectrumGen}. 
We consider the galaxy bispectrum $B_g$ as a function of triangle size for two isosceles shapes. We fix $\mu_{12}=\cos\theta_{12}$ and vary $k$, for two special cases:
\be
\mbox{equilateral:}\quad \mu_{12}={-{1\over2}}, \qquad\qquad \mbox{moderately squeezed:} \quad\mu_{12}={-0.998}~\Rightarrow~ k_3\approx {k \over 16}.
 \ee 
\begin{figure}[!ht]
\centering
\includegraphics[width=0.49\textwidth] {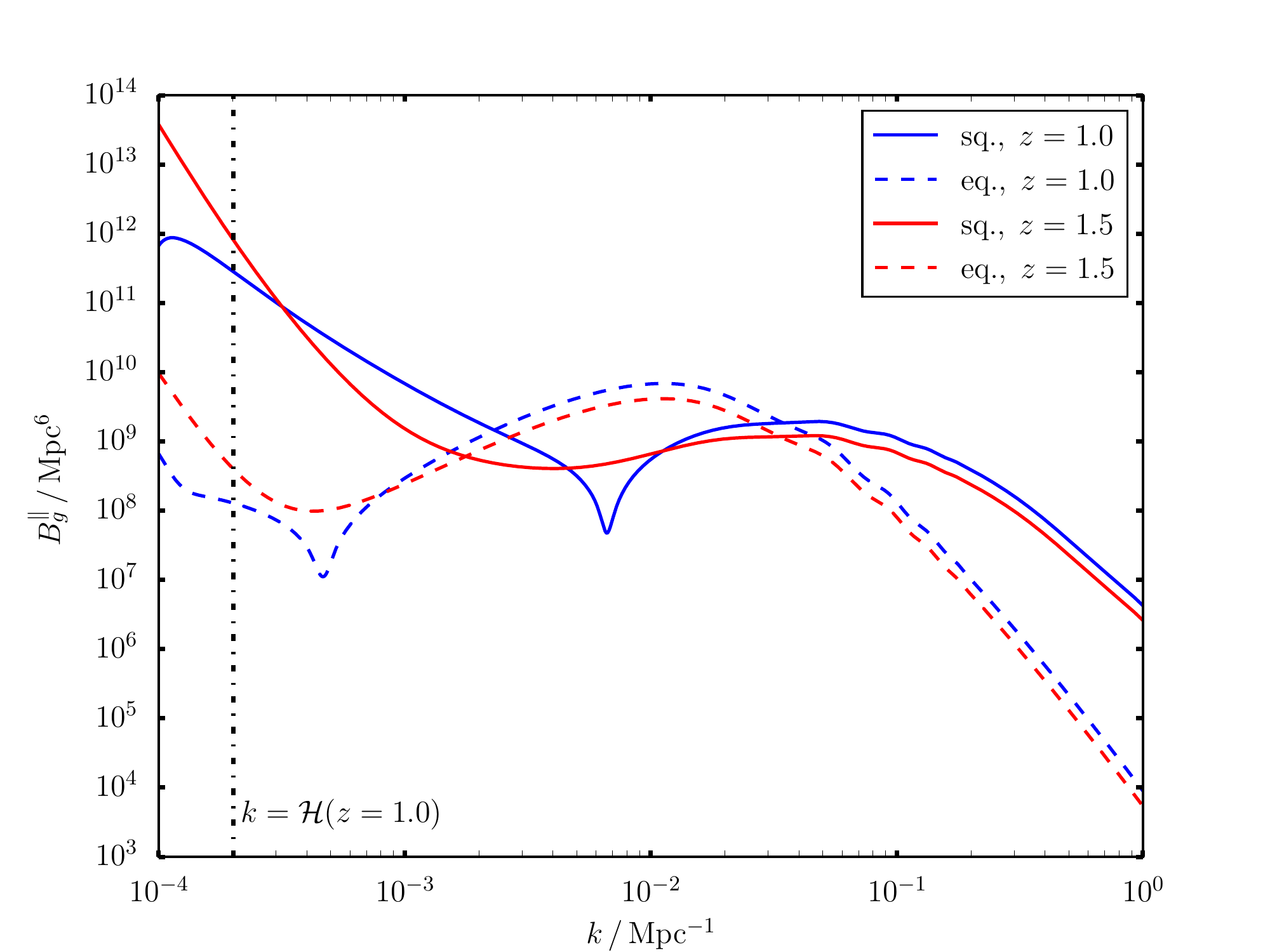}
\includegraphics[width=0.49\textwidth] {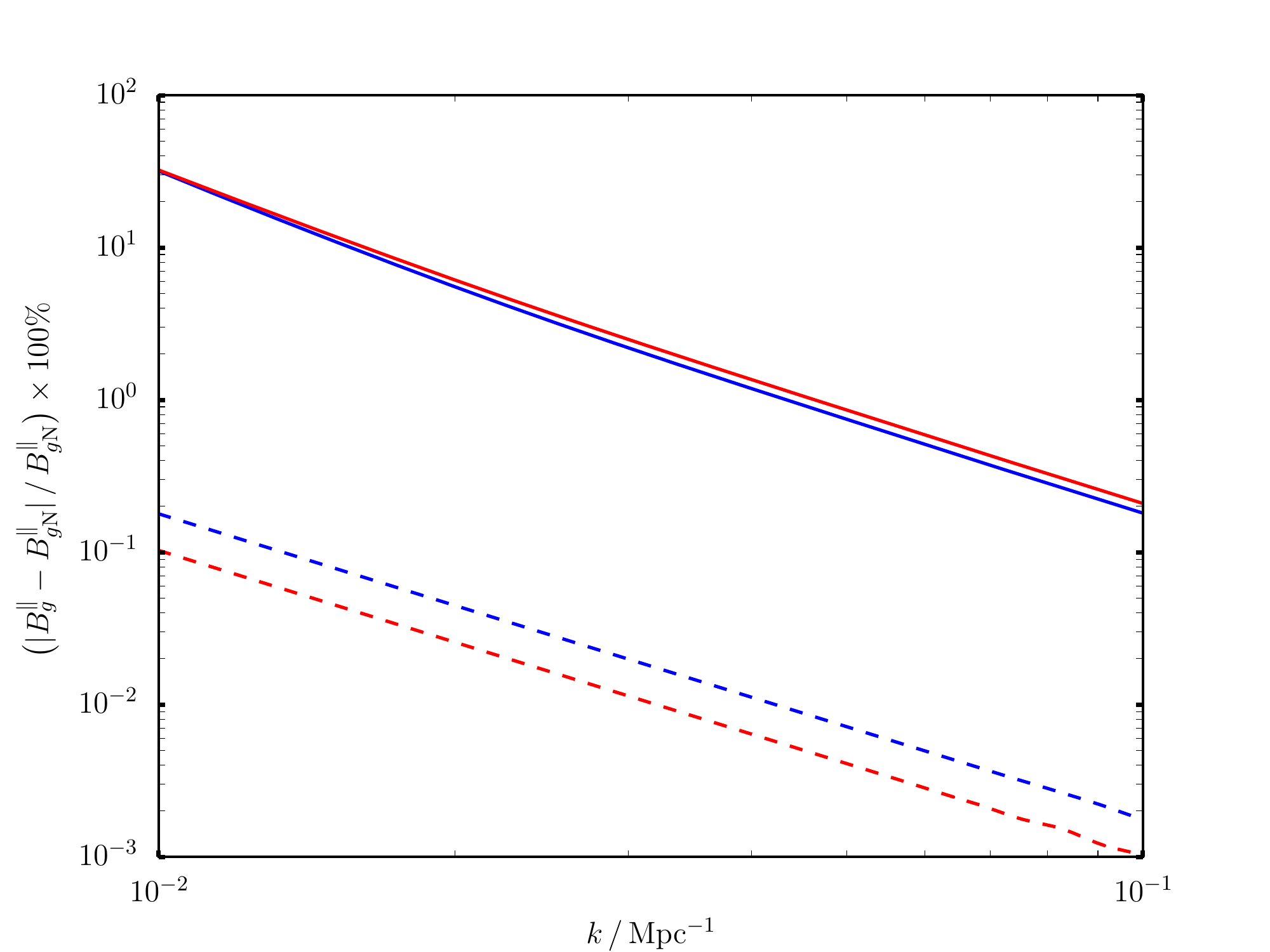} \\
\includegraphics[width=0.49\textwidth] {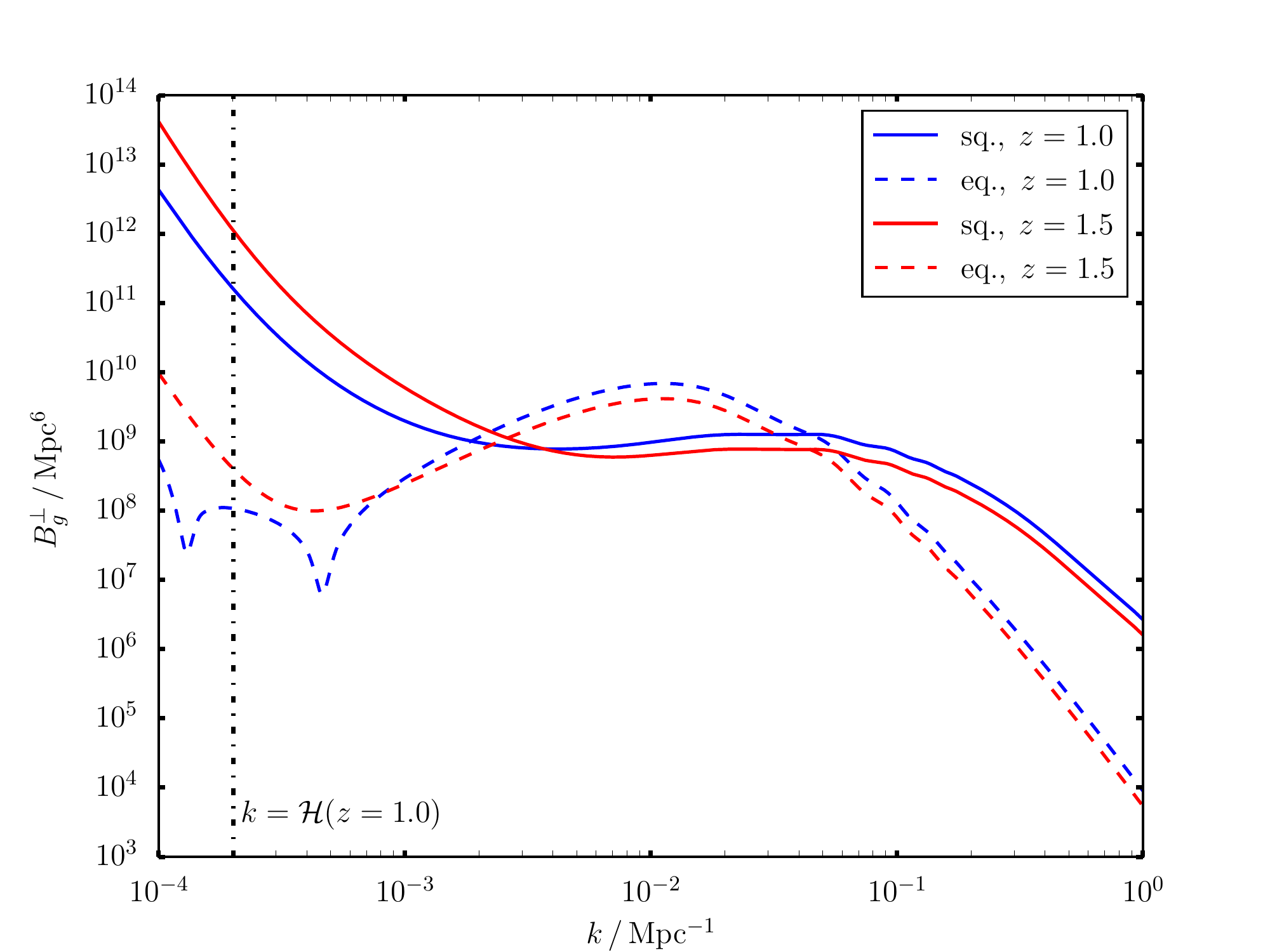}
\includegraphics[width=0.49\textwidth] {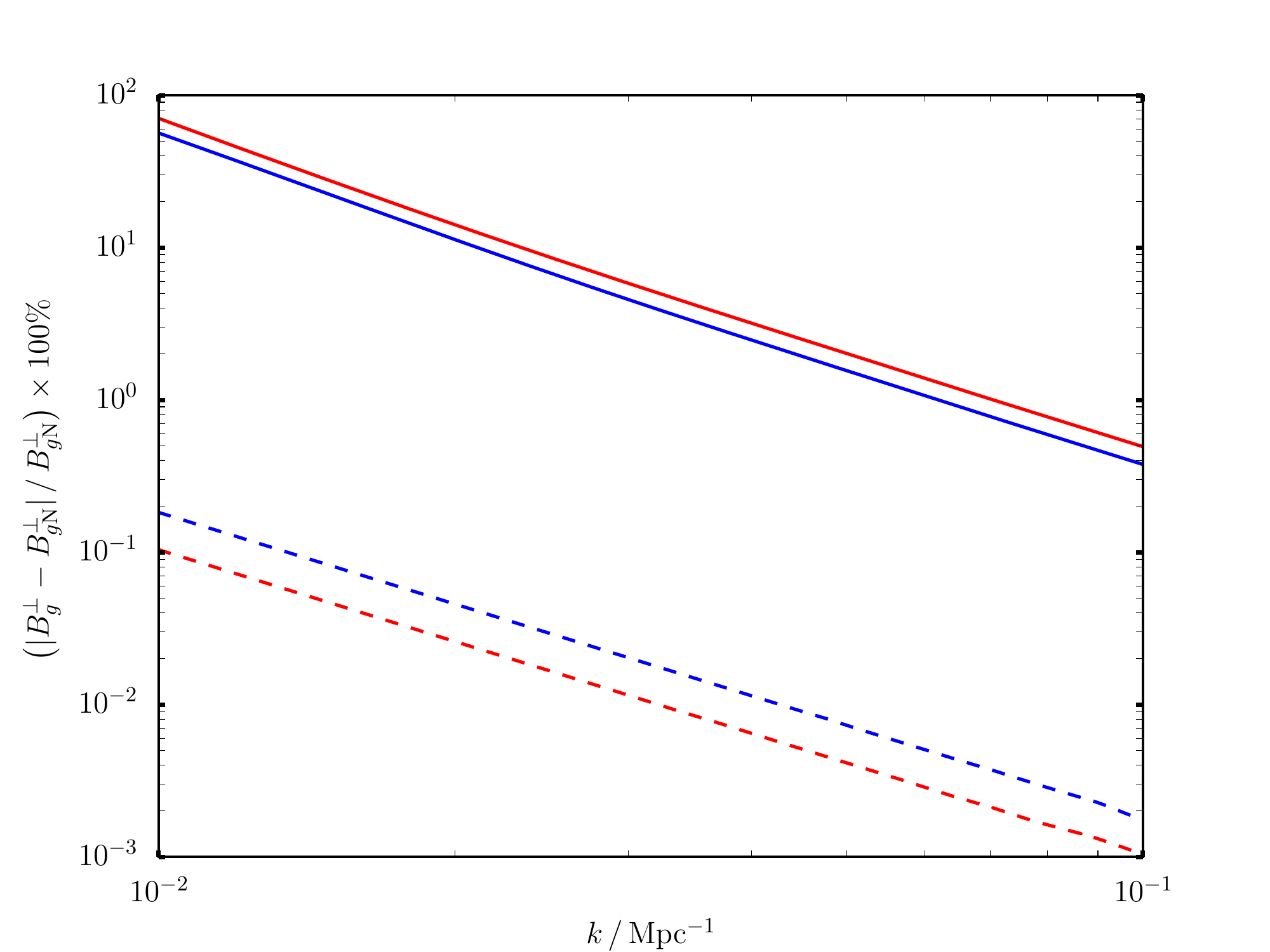}\\
\includegraphics[width=0.49\textwidth] {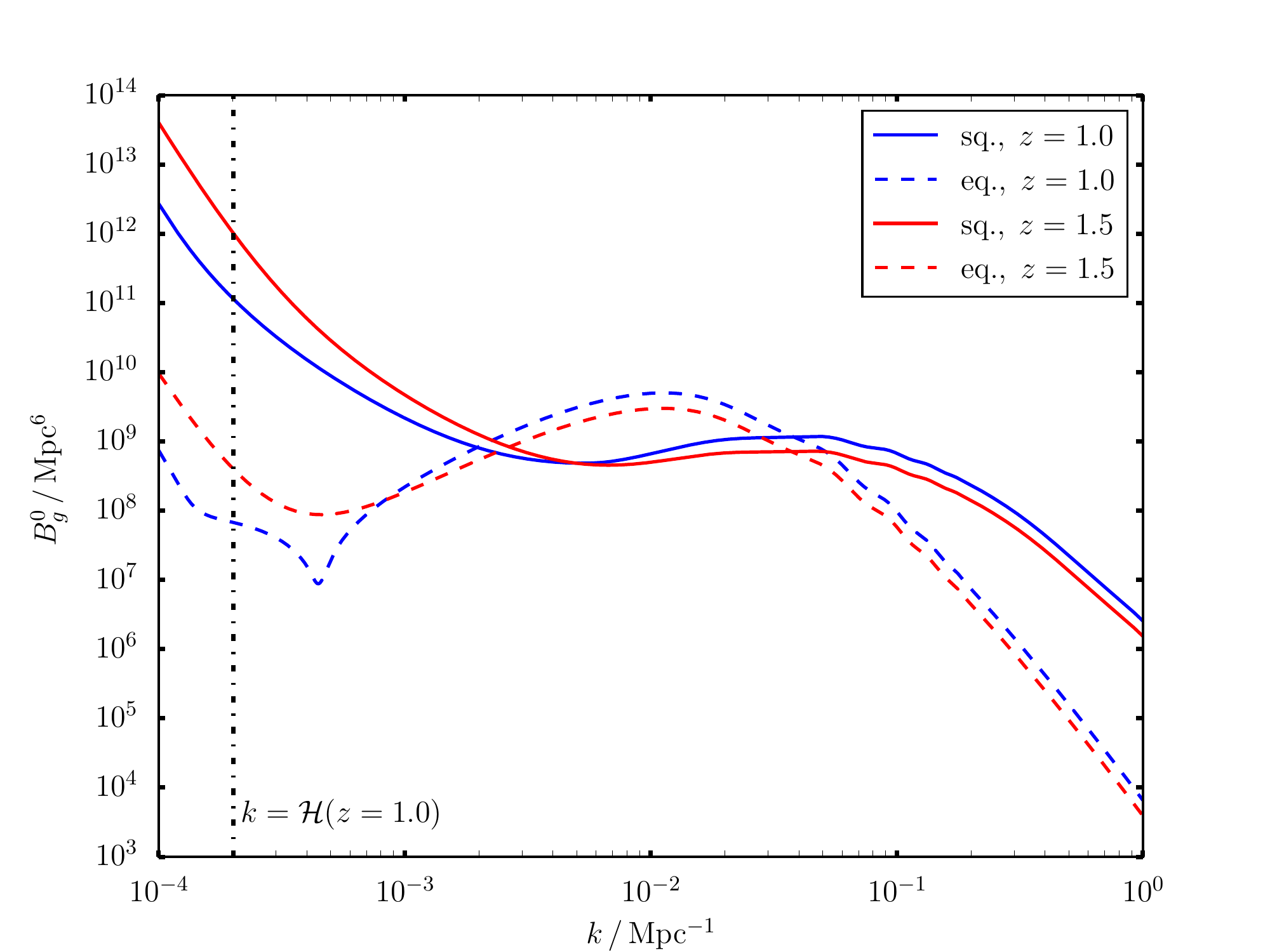}
\includegraphics[width=0.49\textwidth] {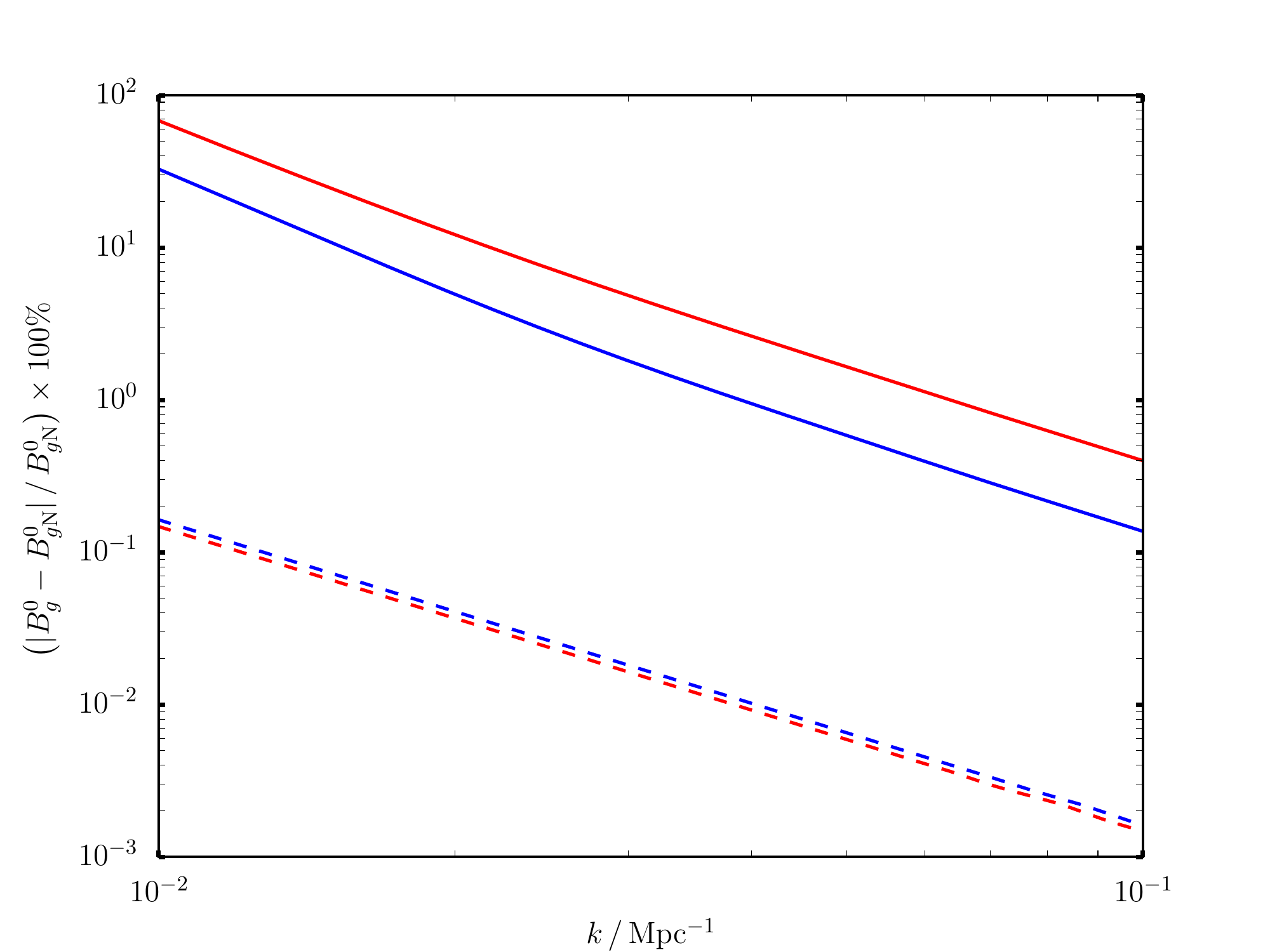}
\caption{{\em Left:} Galaxy bispectrum for moderately squeezed ({$k_3\approx k/16$}, solid) and equilateral ({$k_3=k$}, dashed) shapes, at  {$z=1.0, 1.5$}. From top to bottom: radial, transverse and monopole parts. {\em Right:} {Percentage} difference relative to the Newtonian approximation {for $0.01\leq k\leq0.1$, which includes BAO scales}.}
\label{fig:bispectrum}
\end{figure}
\begin{figure}[!ht]
\centering
\includegraphics[width=0.49\textwidth] {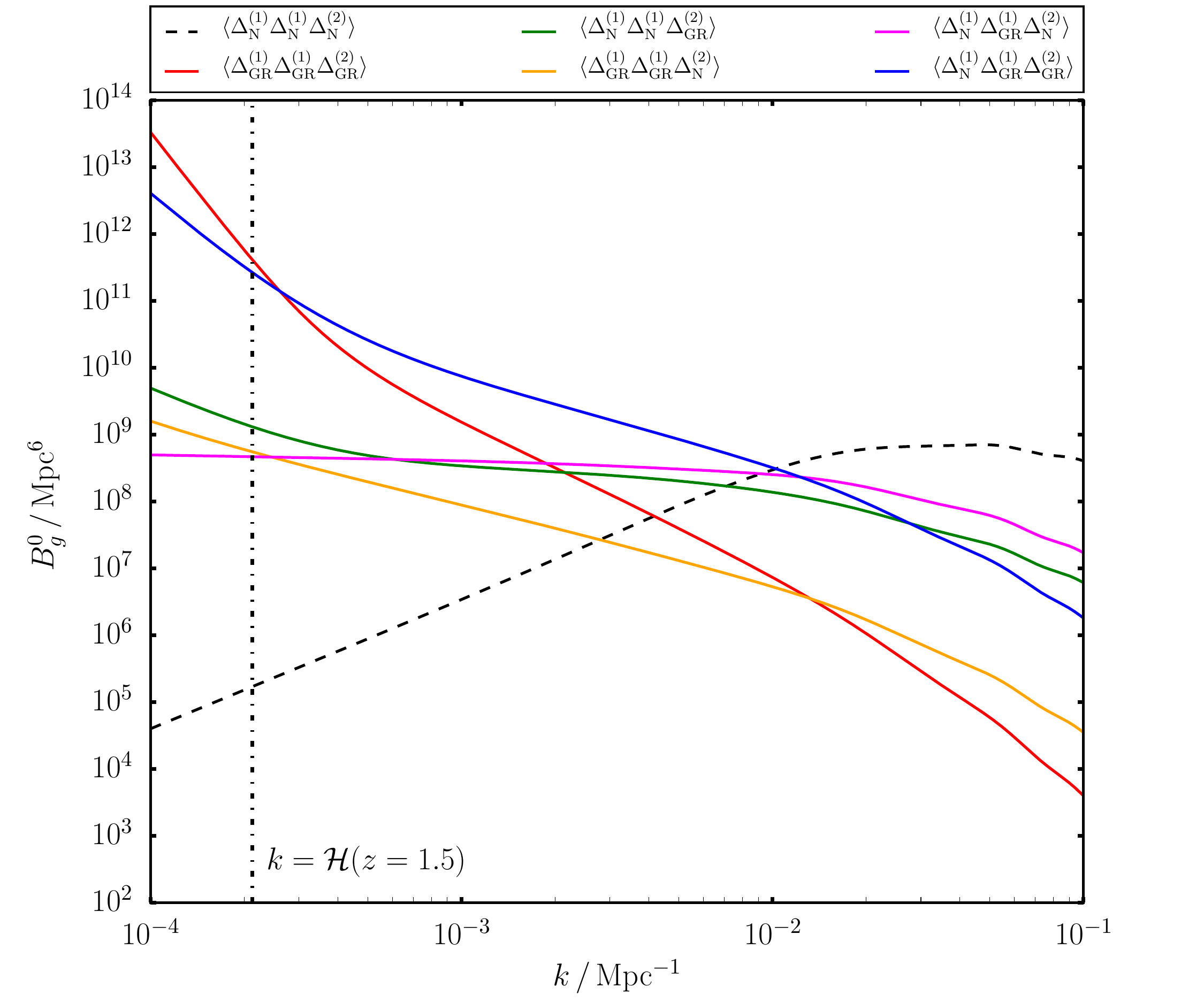}
\includegraphics[width=0.49\textwidth] {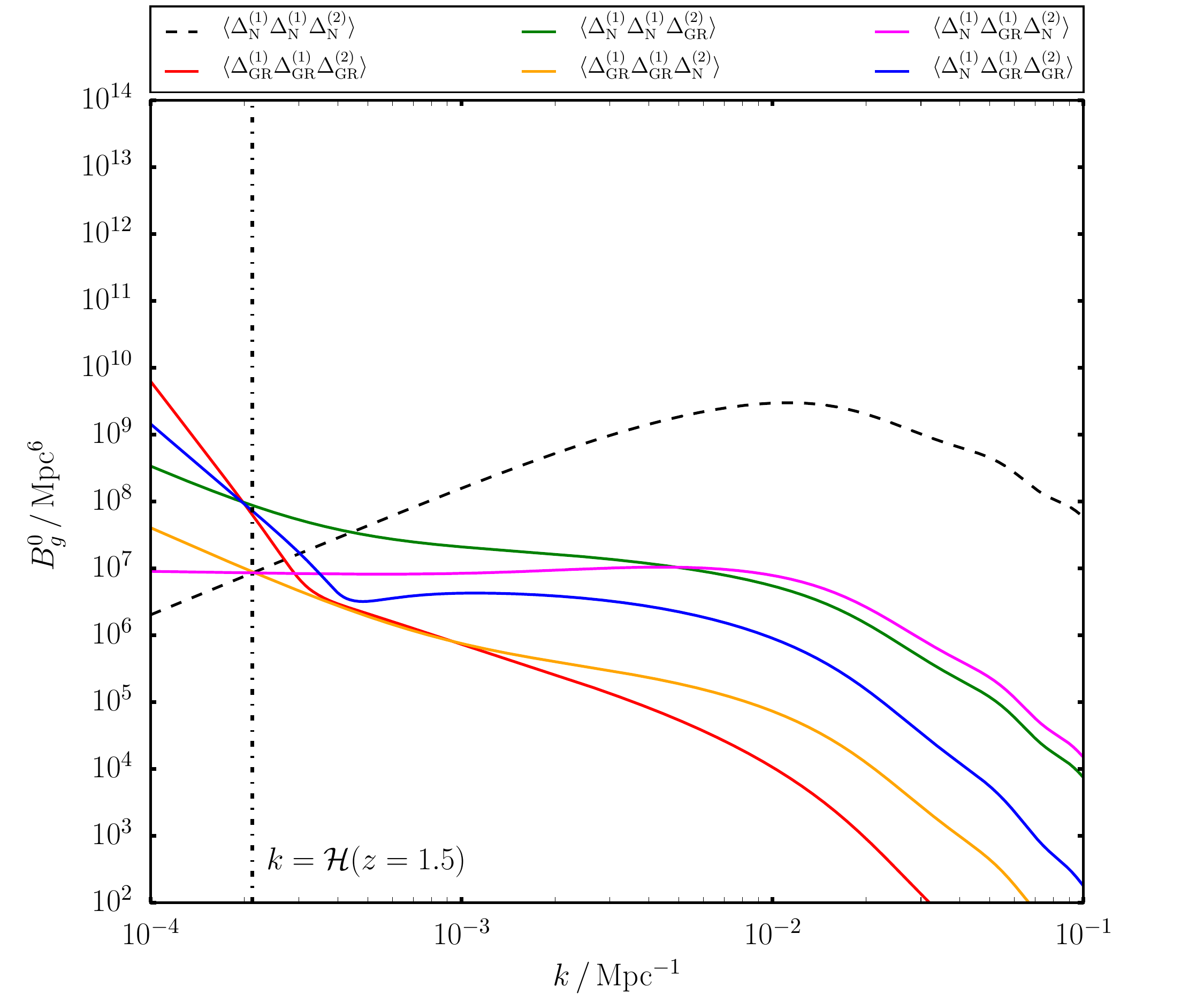} \\
\includegraphics[width=0.49\textwidth] {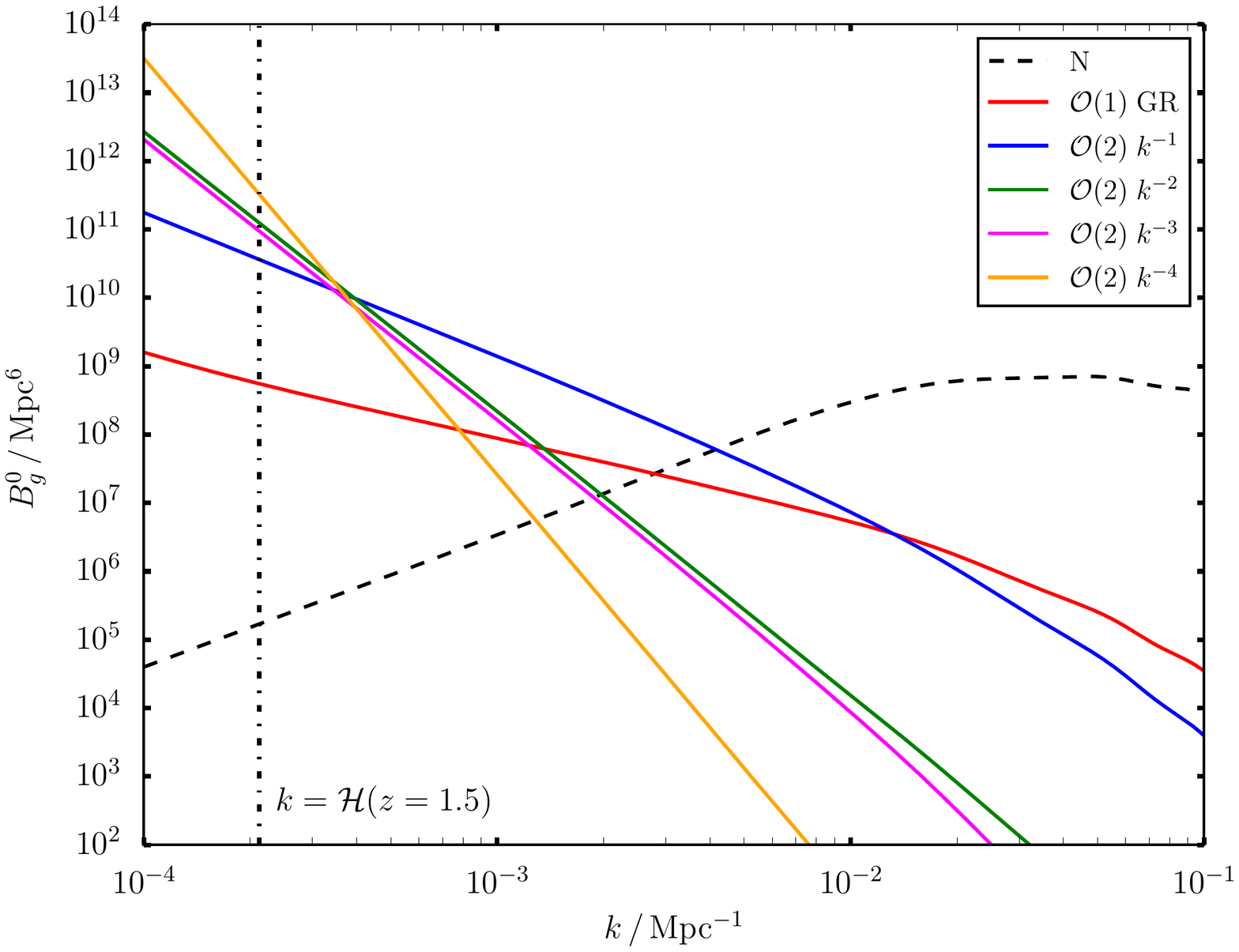}
\includegraphics[width=0.49\textwidth] {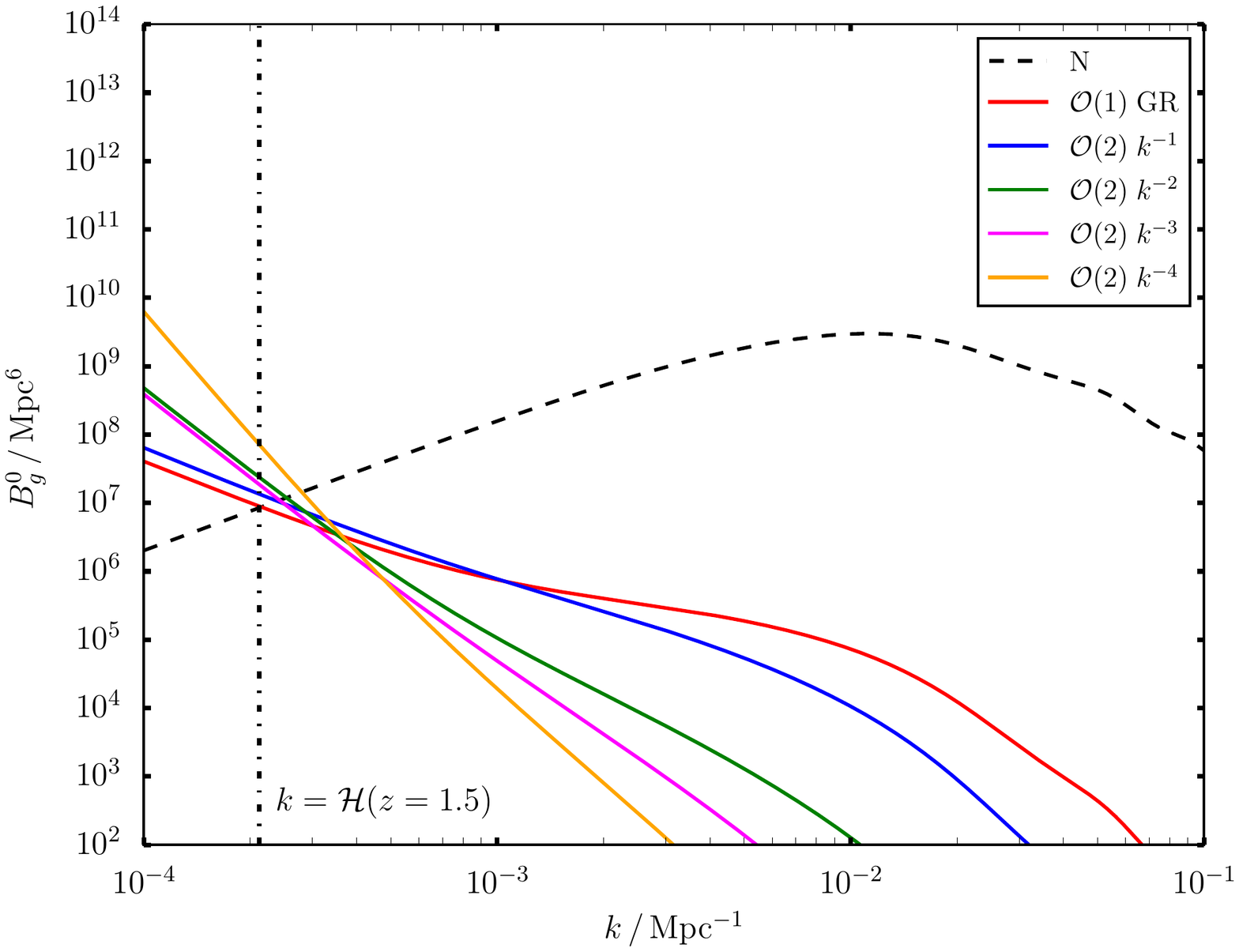}
\caption{{ Contributions to the galaxy bispectrum monopole for the moderately squeezed ({\em left}) and equilateral ({\em right}) shapes of Fig.~\ref{fig:bispectrum}, at $z=1.5$. \\ {\em Top:} The different 3-point correlations that  contribute to the galaxy bispectrum -- purely Newtonian, purely GR and mixed correlations -- as given in \eqref{eq:defbispec}.\\ {\em Bottom:}
The different contributions to the galaxy bispectrum from the first order GR kernel  \eqref{k1} on its own, and then together with the terms in the second order GR kernel \eqref{e8}, split into powers of $k^{-1}$.} }
\label{bparts}
\end{figure}

Figure~\ref{fig:bispectrum}  shows  the radial, transverse and monopole parts of $B_g$, together with the {percentage} correction relative to the Newtonian case without the GR projection effects, on scales $0.01\leq k\leq0.1$, {which includes BAO scales}. In all cases, as expected, the GR corrections become increasingly important on {larger scales}. The squeezed configuration has a larger correction than the equilateral. {For the monopole, the GR correction at equality scales reaches ${\cal O}(30-{70\%})$ at $z\sim 1-1.5$, and then grows larger. Note that when the short modes are equality scale, the long mode is still within the Hubble horizon:
\be
k\sim k_{\rm eq}\quad\Rightarrow\quad k_3\sim {k_{\rm eq}\over 16} \sim  3H_0. 
\ee} 
On the largest scales, our results need to be corrected for wide-angle correlations that are absent in the plane-parallel approximation.   

{It is interesting to identify the various contributions to the galaxy bispectrum monopole in Fig.~\ref{fig:bispectrum}. We do this in two ways, as illustrated
in Fig. \ref{bparts}, for the moderately squeezed (left) and equilateral (right) shapes, at $z=1.5$:\begin{itemize}
\item
In the top panel, we show the contributions from the various 3-point correlations $\big\langle \Delta_g(\k_1) \Delta_g(\k_2)\Delta_g(\k_3)\big\rangle$, as given in \eqref{eq:defbispec}.  

The pure Newtonian correlation gives the standard curve (dashed, black). The 5 solid curves are the correlations with GR corrections: 1 pure GR correlation (red), which dominates on horizon scales, and 4 correlations between GR and Newtonian. It can be seen that 3 of the mixed correlation terms (blue, green, magenta) dominate the GR correction on subhorizon scales. 

For the squeezed case, the dominant correlation is
$\big \langle\Delta^{{{({1})}}}_{g\rm{N}}( \bm{k}_{1})\Delta^{{{({1})}}}_{g\rm{GR}}( \bm{k}_{2})\Delta^{{{({2})}}}_{g\rm{GR}}( \bm{k}_{3})\big\rangle$ (blue). If we omitted the second-order GR projection effects, we would miss this dominant GR contribution to the squeezed galaxy bispectrum.

Note that the correlation with only one GR first-order projection term, i.e., $ \big\langle\Delta^{{{({1})}}}_{g\rm{N}}( \bm{k}_{1})\Delta^{{{({1})}}}_{g\rm{GR}}( \bm{k}_{2})\Delta^{{{({2})}}}_{g\rm{N}}( \bm{k}_{3})\big\rangle$ (magenta), has a constant contribution on super-equality scales. 
\item
In the bottom panel, we show the contributions  from the first-order GR kernel ${\cal K}^{(1)}_{\rm GR}$ , \eqref{k1}, on its own (red), and then together with the terms in the second-order GR kernel  ${\cal K}^{(2)}_{\rm GR}$, \eqref{e8}, split into powers of $k^{-1}$.

The first-order GR correction (red) clearly under-estimates the full GR correction, especially in the squeezed case. 

Amongst the second-order GR corrections in the squeezed case, the $k^{-1}$ term (blue) dominates on ultra-large scales until close to the comoving horizon, $k=\HH$, when the $k^{-n}$, $n=2,3,4$ terms (green, magenta, orange) dominate.

\end{itemize}}
{On scales around equality, we can find a power-law fit for the fractional GR corrections to the Newtonian prediction:
\bea
B_g^{(s)} &=& B_{g{\rm N}}^{(s)}\Big[ 1+\Delta B^{(s)}\Big]\quad s=\, \mbox{radial, transverse, monopole}, \\
 \label{fit1}
 \Delta B^{(s)} &=& \alpha^{(s)}\left({k\over k_{\rm eq}} \right)^{-n}\qquad {0.007\,{\rm Mpc}^{-1} \lesssim k \lesssim 0.07}\,{\rm Mpc}^{-1}.
\eea
We find that $n=2$  {is a good fit for all} $s$ and redshift, and for squeezed and equilateral cases. This shows that the {dominant GR corrections add up to behave as  ${\cal O}(\HH^2/k^2)$  around equality scales.}  The amplitude on equality scales, $\alpha^{(s)}$, varies weakly with $s$ and $z$, but is significantly smaller for equilateral shapes -- see Table~\ref{tab1}.}

\begin{table}[!h]
\centering 
\begin{tabular}{|c|c|c|}
\hline &&\\
  & $~~~\alpha^{(s)}\times 10^2$ for {$z=1,\;1.5~~~$} &~~Triangle shape~~ \\&&  \\ \hline\hline
\multirow{ 2}{*}{$~\Delta B^{\|}~$} & $ 32.8,\;32.2$ & squeezed  \\
& $ .17,\;.096$  & equilateral    \\ \hline
\multirow{ 2}{*}{$~\Delta B^{\bot}~$} & $54.1,\;69.6 $ & squeezed   \\
& $ .17,\;.096 $  & equilateral   \\ \hline
\multirow{ 2}{*}{$~\Delta B^0~$} & $ 33.5,\;69.4 $ & squeezed    \\
& $ .15,\;.14 $ & equilateral    \\ \hline
\end{tabular}
\vspace*{0.2cm}
\caption{{{Percentage} GR corrections at equality, as defined in \eqref{fit1}, for the bispectra in Fig.~\ref{fig:bispectrum}.} }
\label{tab1} 
\end{table}

\newpage

\section{Conclusion}

We considered the local relativistic projection effects on the galaxy bispectrum, up to second order, providing the details behind the results presented in \cite{Umeh:2016nuh}, and generalizing those results to include evolution bias and magnification bias. We transformed the local GR contribution into Fourier space,  to form the kernel $\mathcal{K}^{{{({2})}}}_{\mathrm{GR}}(\bm{k}_{1}, \bm{k}_{2}, \bm{k}_{3})$ given by~\eqref{e8}, {with further details presented in Appendix B, and the $\Gamma_I$ coefficients given in Appendix C}. Once we have this kernel, computing the bispectrum is a relatively straightforward procedure, which allows us to analyse the contribution from  GR effects to the bispectrum. 

{We incorporated a careful treatment of galaxy bias on ultra-large scales, which is essential in order to avoid spurious gauge effects. We assumed a simple local-in-mass-density model of nonlinear bias that neglects tidal effects, leading to the relativistic bias relation \eqref{pog} for the Poisson-gauge galaxy number density contrast.} 

The GR effects can be {significant}, {as illustrated in Fig.~\ref{fig:bispectrum} and Table I, for equilateral and moderately squeezed triangles in the radial, transverse and monopole parts of the bispectrum}. On equality scales {at $z\sim 1-1.5$} they alter the bispectrum monopole {in the moderately squeezed case by {$\sim 30-70\%$}.  {On ultra-large scales,} the bispectrum is dominated by the local GR terms. 

{The contributions to the total GR correction of the monopole are shown in Fig.~\ref{bparts}.
The top panel presents the contributions from the various 3-point correlations given in \eqref{eq:defbispec}.
In the squeezed case, the dominant correlation is
\[
\big \langle\Delta^{{{({1})}}}_{g\rm{N}}( \bm{k}_{1})\Delta^{{{({1})}}}_{g\rm{GR}}( \bm{k}_{2})\Delta^{{{({2})}}}_{g\rm{GR}}( \bm{k}_{3})\big\rangle.
\] 
If we included only the first-order GR projection effects in our analysis, we would miss this dominant GR contribution to the squeezed galaxy bispectrum. The bottom panel breaks down the terms in the second-order GR kernel  ${\cal K}^{(2)}_{\rm GR}$ according to powers of $k^{-1}$.
For the squeezed case, the $k^{-1}$ term dominates on ultra-large scales until close to the comoving horizon, $k=\HH$.}

Our main aim was to highlight the importance of the effects from observations, properly analysed in GR, and to this end, we treated the simplest case, taking the first steps towards a complete analysis. 
We have not included: {
\begin{itemize}
\item
primordial non-Gaussianity;
\item
tidal stress in the galaxy bias;
\item
GR corrections to the $v^{(2)}$, $\Phi^{(2)}$ and $\Psi^{(2)}$ terms that contribute to the projection effects;
\item
the second-order effect of the radiation era on 
initial conditions for sub-equality modes;
 \item
integrated contributions to the projection effects, wide-angle correlations and radial (cross-bin) correlations. 

\end{itemize}
}
The first {three} effects can be incorporated within our Fourier-space analysis using the plane-parallel approximation. The fourth requires numerical integration with a second-order Boltzmann code \cite{Tram:2016cpy}. The last requires one to use the 3-point correlation function, for example through a spherical harmonic decomposition.\footnote{After our paper was completed, \cite{Bertacca:2017dzm} presented a formalism for analysing the 3-point correlation function with all GR effects included, {but without computation of the effects.}}

\[\]
\noindent{\bf Acknowledgments:}\\
 We are especially grateful to Kazuya Koyama for very helpful comments. We thank Tobias Baldauf, Daniele Bertacca, Ruth Durrer, Sabino Matarrese and David Wands for useful discussions and comments. {We also thank an anonymous referee for very useful comments.}
All authors are funded in part by the NRF (South Africa). OU, SJ and RM are also supported by the South African SKA Project. RM {and CC} are also supported by the UK STFC, {Grants ST/N000668/1 (RM) and  ST/P000592/1 (CC)}. 

\newpage

\appendix

\section{Second-order gauge transformation of number density contrast}

At second order, the number density contrasts in Poisson and C gauges are related by  a generalisation of \eqref{eq:lineargaugetrans}, which is given in \cite{Bertacca:2014dra}:
\bea
\delta_{g }^{{{({2})}}}  &=&\delta_{g{\rm C} }^{{{({2})}}}+
(3-b_e) \HH v^{{{({2})}}} +\Big[ (b_e-3) \HH' + {b_e' \HH} + (b_e-3)^2 \HH^2 \Big] \big[v^{{{({1})}}}\big]^2 + (b_e-3)\HH   v^{{{({1})}}}  v^{{{({1})}}\prime}
  \nonumber\\&&{} 
 - \left(b_e-3\right) \HH  \nabla^{-2}\bigg[ v^{{{({1})}}} \nabla^2 {v^{{{({1})}}\prime}}
 - {v^{{{({1})}}\prime}}\nabla^2 v^{{{({1})}}} - 6 \partial_i \Phi^{{{({1})}}} \partial^i v^{{{({1})}}} - 6 \Phi^{{{({1})}}} \nabla^2 v^{{{({1})}}}\bigg]  
    +2(3-b_e)\HH  v^{{{({1})}}} \delta_{g{\rm C}}^{{{({1})}}}  - 2  v^{{{({1})}}} {\delta_{g{\rm C}}^{{{({1})}}\prime}} 
  \nonumber\\ &&{}
   - \frac{1}{2} \partial^i \xi^{{{({1})}}} \left[(3- b_e) \HH \partial_i v^{{{({1})}}}+ 2 \partial_i \delta_{g  {\rm C}}^{{{({1})}}} \right] 
  \nonumber\\ &&{}
 -\frac{1}{2}\left(b_e-3\right) \HH  \nabla^{-2}\bigg[  \partial_i \xi^{{{({1})}}} \partial^i \nabla^2 v^{{{({1})}}} +  \partial_i v^{{{({1})}}} \partial^i \nabla^2 \xi^{{{({1})}}} +2 \partial_i \partial_j \xi^{{{({1})}}} \partial^i \partial^j v^{{{({1})}}}\bigg].
 \label{eq:gGR2}
\eea
{Here $\xi^{(1)}$ is a gauge generator,} and the residual  C-gauge freedom is fixed by imposing $\xi^{{{({1})}}\prime} =2v^{{{({1})}}}$ \cite{Bertacca:2014dra}. 

It follows from the identity
\be
\nabla^2\Big[ \partial_i \xi^{{{({1})}}} \cdot\partial^i v^{{{({1})}}}\Big]=\partial^i v^{{{({1})}}} \cdot\nabla^2\big[\partial_i \xi^{{{({1})}}} \big] + \partial_i \xi^{{{({1})}}}\cdot\nabla^2\big[ \partial^i v^{{{({1})}}}\big]+2\partial_j \partial_i \xi^{{{({1})}}} \cdot\partial^j \partial^i v^{{{({1})}}},
\ee
that the last line of \eqref{eq:gGR2} reduces to $-\left(b_e-3\right) \HH\,\partial_i \xi^{{{({1})}}} \partial^i v^{{{({1})}}}/2$, which cancels the first term on the third line. Thus \eqref{eq:gGR2} may be simplified to
\bea
\delta_{g }^{{{({2})}}}  &=&\delta_{g{\rm C} }^{{{({2})}}}+
(3-b_e) \HH v^{{{({2})}}} +\Big[ (b_e-3) \HH' + {b_e' \HH} + (b_e-3)^2 \HH^2 \Big] \big[v^{{{({1})}}}\big]^2 + (b_e-3)\HH   v^{{{({1})}}}  v^{{{({1})}}\prime}
  \nonumber\\&&{} 
 - \left(b_e-3\right) \HH  \nabla^{-2}\bigg[ v^{{{({1})}}} \nabla^2 {v^{{{({1})}}\prime}}
 - {v^{{{({1})}}\prime}}\nabla^2 v^{{{({1})}}} - 6 \partial_i \Phi^{{{({1})}}} \partial^i v^{{{({1})}}} - 6 \Phi^{{{({1})}}} \nabla^2 v^{{{({1})}}}\bigg]  
    +2(3-b_e)\HH  v^{{{({1})}}} \delta_{g{\rm C}}^{{{({1})}}}  - 2  v^{{{({1})}}} {\delta_{g{\rm C}}^{{{({1})}}\prime}} 
  \nonumber\\ &&{}
-\big[\partial_i\delta_{gC}^{(1)}\big]\partial^i\xi^{(1)}.
 \label{pt2}
\eea

By the continuity equation, given in \eqref{pc}, the gauge fixing condition $\xi^{{{({1})}}\prime} =2v^{{{({1})}}}$ implies that
\be\label{xi}
\partial^i\xi^{(1)}=-2\nabla^{-2}\partial^i\delta_{m{\rm C}}^{(1)}.
\ee
Using this, the relation
\eqref{15a} between C- and T-gauge number density contrasts becomes
\be\label{gct2}
\delta_{g{\rm C} }^{{{({2})}}}-\big[\partial_i\delta_{gC}^{(1)}\big]\partial^i\xi^{(1)}= \delta_{g{\rm T} }^{{{({2})}}}.
\ee
Then it follows from  \eqref{pt2} and \eqref{gct2} that \eqref{eq:gGR2}  can be rewritten as the second-order map from the Poisson-gauge $\delta_g$ to the T-gauge $\delta_{g{\rm T} }$:
\bea
\delta_{g }^{{{({2})}}}  &=&\delta_{g{\rm T} }^{{{({2})}}}+
(3-b_e) \HH v^{{{({2})}}} +\Big[ (b_e-3) \HH' + {b_e' \HH} + (b_e-3)^2 \HH^2 \Big] \big[v^{{{({1})}}}\big]^2 + (b_e-3)\HH   v^{{{({1})}}}  v^{{{({1})}}\prime}
  \nonumber\\&&{} 
 - \left(b_e-3\right) \HH  \nabla^{-2}\bigg[ v^{{{({1})}}} \nabla^2 {v^{{{({1})}}\prime}}
 - {v^{{{({1})}}\prime}}\nabla^2 v^{{{({1})}}} - 6 \partial_i \Phi^{{{({1})}}} \partial^i v^{{{({1})}}} - 6 \Phi^{{{({1})}}} \nabla^2 v^{{{({1})}}}\bigg] 
  \nonumber\\&&{} 
    +2(3-b_e)\HH  v^{{{({1})}}} \delta_{g{\rm T}}^{{{({1})}}}  - 2  v^{{{({1})}}} {\delta_{g{\rm T}}^{{{({1})}}\prime}} .
 \label{dgt2a}
\eea
This is \eqref{dgt2}.

\newpage
\section{Expansion of perturbed variables in Fourier space} 

We express all variables in terms of the {T-gauge} matter density contrast, $\delta({\k})$.
For the gravitational and velocity potentials,  \eqref{phi}, \eqref{vm} and \eqref{1} give 
\begin{eqnarray}\label{vkpk}
\HH \,v^{{{({1})}}}(\bm{k}) =f \frac{\HH^2}{k^2}\,\delta^{{{({1})}}}(\bm{k}), \qquad \qquad
 \Phi^{{{({1})}}}(\bm{k})= -\frac{3}{2}\Omega_m \frac{\HH^2}{k^2}\,\delta^{{{({1})}}}(\bm{k}). 
\end{eqnarray} 
The growth rate and growth suppression factor in $\Lambda$CDM obey
\begin{equation} \label{e18}
\frac{f'}{\mathcal{H}}={1\over2}\big(3\Omega_m-4 \big)f-f^2 +{3\over2}\Omega_m,\qquad  \qquad  
{1\over\mathcal{H}}\frac{g'}{g} = f-1.
\end{equation}
The galaxy number density contrast in Fourier space is expanded using \eqref{sb}, \eqref{eq:lineargaugetrans}:
\begin{equation}\label{dgk}
\delta_{g }^{{{({1})}}}
=b_1\delta^{{{({1})}}}+\left(3- b_e \right)\HH v^{{{({1})}}}.
\end{equation}
The evolution of the velocity potential follows from the Euler equation as
\begin{eqnarray}
v^{{{({1})}}\prime}=- \HH v^{{{({1})}}} - \Phi^{{{({1})}}} .\label{euler}
\end{eqnarray}
The time derivative of the galaxy number density contrast follows from \eqref{dgk} and \eqref{euler} as
\be 
\label{eq:nofprterm}
\delta_g^{{{({1})}}\prime}= \big({b_1'}+b_1 f \HH\big) \delta^{{{({1})}}} +\big[(3-b_e) \big({\HH'}-{\HH^2} \big)  - {b_e'}{\HH}\big] v^{{{({1})}}} -(3-b_e)\HH \Phi^{{{({1})}}} .
\ee

 At second order, a typical term such as $v^{{{({1})}}}({\x})\delta_g^{{{({1})}}}({\x})$ can be expressed as:  
\begin{eqnarray}
v^{{{({1})}}}({\x})\delta_g^{{{({1})}}}({\x})&=& \int \frac{\d^{3} k}{(2 \pi)^{3}} e^{{\rm i} {\bm{k} \cdot \bm{x}}} {\left[v^{{{({1})}}}\delta_g^{{{({1})}}}\right]}({\k}),\\
\left[ v^{{{({1})}}}\delta_g^{{{({1})}}}\right]({\k})&=& \int \d^{3} x\,  e^{-{\rm i} {\bm{k} \cdot \bm{x}}}v^{{{({1})}}}({\x})\delta_g^{{{({1})}}}({\x})
\nonumber\\ 
&=&\frac{1}{2}\int \d^{3} x \int \frac{\d^{3} k_1}{(2 \pi)^{3}}\frac{\d^{3} k_2}{(2 \pi)^{3}} \left[v^{{{({1})}}}({\k}_1) \delta_g^{{{({1})}}}({\k}_2)+ v^{{{({1})}}}({\k}_2) \delta_g^{{{({1})}}}({\k}_1)\right] e^{-{\rm i} {\bm{k} \cdot \bm{x}}} e^{{\rm i} {\bm{k}_1 \cdot \bm{x}}} e^{{\rm i} {\bm{k}_2 \cdot \bm{x}}} \nonumber\\
&=&\frac{1}{2}\int \frac{\d^{3} k_1}{(2 \pi)^{3}}\frac{\d^{3} k_2}{(2 \pi)^{3}} \left[v^{{{({1})}}}({\k}_1) \delta_g^{{{({1})}}}({\k}_2)+ v^{{{({1})}}}({\k}_2) \delta_g^{{{({1})}}}({\k}_1)\right] (2\pi)^3\delta^{D}\left({\k}_1 + {\k}_2-{\k}\right),
\end{eqnarray}
where we used \eqref{eq:STF1} and the definition of
the Dirac delta function in three dimensions.
Then we express the perturbative variables in terms of $\delta^{(1)}$, using \eqref{vkpk} and \eqref{dgk}:
\begin{eqnarray}
	v^{{{({1})}}}({\k}_1) \delta_g^{{{({1})}}}({\k}_2)+ v^{{{({1})}}}({\k}_2) \delta_g^{{{({1})}}}({\k}_1) &=&  \bigg[b_1f{\HH} \bigg( \frac{1}{k^2_1}+\frac{1}{k^2_2} \bigg) 
	+ {2f^2 \left(3 -b_e\right)\HH^3\, \frac{1}{k^2_1k_2^2}\bigg]\delta^{(1)}({\k}_1)\delta^{(1)}({\k}_2).}
\end{eqnarray}
This leads to 
\begin{eqnarray}
{\left[ v^{{{({1})}}}\delta_g^{{{({1})}}}\right]}({\k})	&=&\int \frac{\d^{3} k_1}{(2 \pi)^{3}} \frac{\d^{3} k_2}{(2 \pi)^{3}}\mathcal{F}{\left[v^{{{({1})}}}\delta_g^{{{({1})}}}\right]}\delta^{(1)}({\k}_1)\delta^{(1)}({\k}_2)
	(2\pi)^3 {\delta^D}\left({\k}_1 + {\k}_2-{\k}\right),
\end{eqnarray}
where the kernel is
\begin{eqnarray}
\mathcal{F}\left[v^{{{({1})}}}({\x})\delta_g^{{{({1})}}}({\x})\right]&=& f\HH\,{\left[b_1\left({k^2_1}+{k^2_2}\right)  + 2\left(3 -b_e\right)f\HH^2\right] \over 2k_1^2k_2^2}.
\end{eqnarray}
	
Table \ref{table1} gives the Fourier kernels for all second-order terms in $\Delta_g^{(2)}$.

\pagebreak
\FloatBarrier
\begin{table}[!t] 
\centering 
\caption{{Fourier transform kernel and coefficient of each term of \eqref{eq:SecondorderNewtonian} and \eqref{eq:secondorderGRA}, ordered according to their $k$-dependence. N denotes a Newtonian term ($k^{0}$), $\Gamma_{1}$ is for $k^{-4}$, $\Gamma_{2}$ is for $k^{-3}$, $\Gamma_{3}$ to $\Gamma_{8}$ are for $k^{-2}$ and $\Gamma_{9}$ to $\Gamma_{14}$ are for $k^{-1}$. For convenience, the superscript (1) is dropped from first-order variables $\delta^{(1)},v^{(1)},\Phi^{(1)}$.}} \label{table1} 
\vspace*{0.2cm}
\begin{tabular}{p{0.35\textwidth-2\tabcolsep - 1.25\arrayrulewidth}
			c 
			c
			c} 
\hline \\
Term & $~~\Gamma~~$ & Fourier kernel ${\cal F}$ & Coefficient \\ \\ \hline \hline \\
$\delta^{(2)}$ & N & $F_{2}(\bm{k}_{1},\bm{k}_{2})$ & $b_{1}$ \\ \\
$\p_{\parallel}^{2}v^{(2)}$ & N & $f^{2}\HH \mu_{3}^{2}G_{2}(\bm{k}_{1},\bm{k}_{2})$ & $-1/\HH$ \\ \\
$\delta\p_{\parallel}^{2}v$ & N & $-f\HH \big(\mu_{1}^{2} + \mu_{2}^{2}\big)/2 $ & $-2b_{1}/\HH$ \\ \\
$\p_{\parallel}v\p_{\parallel}\delta$ & N\ & $-f\HH {{\mu_{1}\mu_{2}\big(k_{1}^{2} + k_{2}^{2}\big)}/{\big(2k_{1}k_{2}\big)}}$& $-2b_{1}/\HH$\\ \\
$\p_{\parallel}v\p_{\parallel}^{3}v $  & N & $f^{2} \HH^2{{\big(\mu_{1}\mu_{2}^{3}k_{2}^{2} + \mu_{2}\mu_{1}^{3}k_{1}^{2}\big)}/{\big(k_{1}k_{2}\big)}}$ & $ {2}/{\HH^{2}}$  \\ \\
$\big[\p_{\parallel}^{2}v\big]^{2}$  & N &$f^{2}\HH^2 \,\mu_{1}^{2}\mu_{2}^{2}$ & ${2}/{\HH^{2}}$ \\ \\ 
 \hline \\
$\big[\Phi\big]^{2}$ & $\Gamma_{1}$ & ${9}\Omega_{m}^{2}\HH^{4}/{\big(4k_{1}^{2}k_{2}^{2}\big)}$ & $\mathcal{A}$ \\ \\
$\Phi v$ & $\Gamma_{1}$ & $-{3}\Omega_{m}\HH^{3}f/{\big(2k_{1}^{2}k_{2}^{2}\big)}$  & $\mathcal{C}$ \\ \\
$\nabla^{-2}(v\nabla^{2}v' - v'\,\nabla^{2}v - 6\partial_{i}\Phi\partial^{i}v- 6\Phi\nabla^{2}v)$ & $\Gamma_{1}$ & ${9}\Omega_{m}\HH^{3}f/{\big(2k_{1}^{2}k_{2}^{2}\big)}$ & $ (3-b_{e})\HH$ \\ \\
$vv' $ & $\Gamma_{1}$  & ${ f\HH^{3}\big(3\Omega_{m}-2f\big)/{\big(2k_{1}^{2}k_{2}^{2}\big)} }$ & $ (b_{e}-3)\HH$ \\ \\
$\big[v\big]^{2}$ & $\Gamma_{1}$ & $f^{2} \HH^{2}/{\big(k_{1}^{2}k_{2}^{2}\big)}$ & $(b_{e}-3)^{2}\HH^2+{b_{e}'\HH} +(b_{e}-3){\HH'} $ \\ \\
 \hline \\
$v\p_{\parallel}v $ & $\Gamma_{2}$ & $ \mathrm{i}\,f^{2}\HH^{2}
{\big(\mu_{1}k_{1} + \mu_{2}k_{2}\big)}/{\big(2k_{1}^{2}k_{2}^{2}\big)}$ &  $\mathcal{B}$ \\ \\
$\Phi\p_{\parallel}v$ & $\Gamma_{2}$ & $-3{\rm i}\,f\Omega_{m}\HH^{3}\,{\big(\mu_{1}k_{1} + \mu_{2}k_{2}\big)}/{\big(4k_{1}^{2}k_{2}^{2}\big)}$ & $\mathcal{D}$ \\ \\ 
$\Phi\p_{\parallel}\Phi$ & $\Gamma_{2}$ & $9{\rm i}\,\Omega_{m}^{2}\HH^{4}{\big(\mu_{1}k_{1} + \mu_{2}k_{2}\big)}/{\big(8k_{1}^{2}k_{2}^{2}\big)}$ & ${2}(f-2+2\Q)/{\HH}$  \\ \\
 \hline \\
$\Psi^{(2)}=\Phi^{(2)}$ & $\Gamma_{3}$ & $-{3}\Omega_{m}{\HH^{2}}F_{2}(\bm{k}_{1},\bm{k}_{2})/{\big(2k_{3}^{2}\big)}$ & $4\Q-1-b_{e}+R$\\ \\
${\Phi^{(2)\prime}}$  & $\Gamma_{3}$ & $-{3}\Omega_{m}{\HH^{3}}(2f-1)F_{2}(\bm{k}_{1},\bm{k}_{2})/{\big(2k_{3}^{2}\big)}$ & ${1}/{\HH}$ \\ \\
 \hline \\
$v^{(2)}$ & $\Gamma_{4}$ & $f{\HH}G_{2}(\bm{k}_{1},\bm{k}_{2})/{k_{3}^{2}}$ & $(3-b_{e})\HH$ \\ \\
 \hline \\
$\big[\p_{\parallel}v\big]^{2} $ & $\Gamma_{5}$ & $-f^{2}\HH^{2}{\mu_{1}\mu_{2}}/{\big(k_{1}k_{2}\big)}$ & $\mathcal{E}$ \\ \\
$\p_{\parallel}v\p_{\parallel}\Phi$ & $\Gamma_{5}$ & ${3}f\Omega_{m}\HH^{3}{\mu_{1}\mu_{2}/{\big(2k_{1}k_{2}\big)}}$ & ${2}(2-f-2\Q)/{\HH}$ \\ \\
 \hline \\
$\p_{i}v\,\p^{i}v$ & $\Gamma_{6}$ & $-f^{2}\HH^{2}\,{\bm{k}_1\cdot \bm{k}_2}/{\big(k_{1}^{2}k_{2}^{2}\big)}$  & $ b_{e} -1- 2\Q - R$ \\ \\
$\p_{i}v\p^{i}\Phi$  &$\Gamma_{6}$ & $3f\Omega_{m}\HH^{3} \,{\bm{k}_{1}\cdot \bm{k}_{2}}/{\big(2k_{1}^{2}k_{2}^{2}\big)} $ & ${2}/{\HH}$  \\ \\
 \hline \\
\end{tabular}
\end{table}

\FloatBarrier
\begin{table}[!t] 
\centering  
\vspace*{0.2cm}
\begin{tabular}{p{0.13\textwidth-2\tabcolsep - 1.25\arrayrulewidth}
		        p{0.05\textwidth-2\tabcolsep - 1.25\arrayrulewidth} 
		        c 
		        c}  
\hline\\
$\Phi\delta$ & $\Gamma_{7}$ &$-{3}\Omega_{m}\HH^{2}{\big(k_{1}^{2}+k_{2}^{2}\big)}/{\big(4k_{1}^{2}k_{2}^{2}\big)} $& $2b_{1}\big(f-2-b_{e}+4\Q+R\big) -S$  \\ \\ 
$\Phi\delta'$& $\Gamma_{7}$ & $-{3}f\Omega_{m}\HH^{3}{\big(k_{1}^{2}+k_{2}^{2}\big)}/{\big(4k_{1}^{2}k_{2}^{2}\big)} $ & $-{2}b_{1}/{\HH}$ \\ \\
$v\delta$& $\Gamma_{7}$ & $f\HH{\big(k_{1}^{2}+k_{2}^{2}\big)}/{\big(2k_{1}^{2}k_{2}^{2}\big)} $  & $b_1'+2b_{1}(3-b_{e})\HH$ \\ \\
$v\delta'$ & $\Gamma_{7}$ & $f^2\HH^2{\big(k_{1}^{2}+k_{2}^{2}\big)}/{\big(2k_{1}^{2}k_{2}^{2}\big)} $ & $-2b_1$ \\ \\
 \hline \\
$\Phi \p^{2}_{\parallel}v$ & $\Gamma_{8}$ & ${3}f\Omega_{m}\HH^{3}{\big(\mu_{1}^{2}k_{1}^{2} + \mu_{2}^{2}k_{2}^{2}\big)}/{\big(4k_{1}^{2}k_{2}^{2}\big)}$ &${2}\big({1}-2f+2b_{e}-{6}\Q-2R-{\HH'}/{\HH^{2}}\big)/{\HH}$ \\ \\
$\Phi\p_{\parallel}^{2}\Phi $ & $\Gamma_{8}$ & $- {9}\Omega_{m}^2\HH^{4}{\big(\mu_{1}^{2}k_{1}^{2} + \mu_{2}^{2}k_{2}^{2}\big)}/{\big(4k_{1}^{2}k_{2}^{2}\big)}$ & $-{2}/{\HH^{2}}$  \\ \\
$v\p_{\parallel}^{2}v$ &$\Gamma_{8}$ & $-f^2\HH^{3}{\big(\mu_{1}^{2}k_{1}^{2} + \mu_{2}^{2}k_{2}^{2}\big)}/{\big(2k_{1}^{2}k_{2}^{2}\big)}$ & ${2}(b_{e}-3)/{\HH}$ \\ \\
 \hline \\
$\Phi\p_{\parallel}\delta$ &$\Gamma_{9}$ & $-3{\rm i}\,\Omega_{m}\HH^2 {\big(\mu_{1}k_{1}^{3} + \mu_{2}k_{2}^{3}\big)}/{\big(4k_{1}^{2}k_{2}^{2}\big)}$& ${2}b_{1}/{\HH}$ \\ \\
 \hline \\
$\p_{i}v\p_{\parallel}\p^{i}v $ & $\Gamma_{10}$ &$ -\mathrm{i}\,f^2\HH^2 \bm{k}_{1}\cdot\bm{k}_{2}{\big(\mu_{1}k_{1} + \mu_{2}k_{2}\big)}/{\big(2k_{1}^{2}k_{2}^{2}\big)}$ & $-{4}/{\HH}$  \\ \\
 \hline \\
$\delta'\p_{\parallel}v$ & $\Gamma_{11}$  & $\mathrm{i}\,f^{2}\HH^{2}{\big(\mu_{1}k_{2} + \mu_{2}k_{1}\big)}/{\big(2k_{1}k_{2}\big)}$ & ${2}b_{1}/{\HH}$ \\ \\
$\delta\p_{\parallel}v $ & $\Gamma_{11}$ & $\mathrm{i}\,f\HH{\big(\mu_{1}k_{2} + \mu_{2}k_{1}\big)}/{\big(2k_{1}k_{2}\big)}$ & $2b_{1}\big(b_{e}-2\Q-R \big)+ S$  \\ \\
 \hline \\
$\Phi\p_{\parallel}^{3}v$ & $\Gamma_{12}$ & $3{\rm i}\,f\Omega_{m}\HH^3 {\big(\mu_{1}^{3}k_{1}^{3} + \mu_{2}^{3}k_{2}^{3}\big)}/{\big(4k_{1}^{2}k_{2}^{2}\big)}$ & $-{2}/{\HH^{2}}$ \\ \\
 \hline \\
$\p_{\parallel}v\p^{2}_{\parallel}v$ & $\Gamma_{13}$ & $-\mathrm{i}\,f^{2}\HH^2 {{\big(\mu_{1}\mu_{2}^{2}k_{2} +\mu_{2}\mu_{1}^{2}k_{1}\big) }/{\big(2k_{1}k_{2}\big)}}$ & ${2}\big(3-2b_{e}+{4}\Q+2R+{\HH'}/{\HH^{2}}\big)/{\HH}$ \\ \\
$\p_{\parallel}v\p^{2}_{\parallel}\Phi$ & $\Gamma_{13}$ & $3{\rm i}\,f\Omega_{m}\HH^3{{\big(\mu_{1}\mu_{2}^{2}k_{2} +\mu_{2}\mu_{1}^{2}k_{1}\big) }/{\big(4k_{1}k_{2}\big)}}$ & ${2}/{\HH^{2}}$ \\ \\
 \hline \\
$\p_{\parallel}v^{(2)}$& $\Gamma_{14}$ & $\mathrm{i}\,f\HH \,{\mu_{3}}G_{2}(\bm{k}_{1},\bm{k}_{2})/{k_{3}}$ & $b_{e}-2Q-R$ \\ \\
\hline
\end{tabular}
\end{table}
where {${\cal A}, {\cal B}, {\cal C}, {\cal D},{\cal E}$ are given by \eqref{bca}--\eqref{bce}, and}
\be
R\equiv \frac{2(1-\Q)}{\chi\HH}+\frac{\HH'}{\HH^{2}}, \qquad
S\equiv 4\bigg(2-\frac{1}{\chi\HH}\bigg)\frac{\p b_{1}}{\p \ln{\bar{L}}}.
\ee 

Note that the kernels for quadratic terms in Table \ref{table1} can be obtained from an algorithm. Consider a term such as
\be
{\rm D}^n X\, {\rm D}^m Y,
\ee
where ${\rm D}=\partial_i$ or $\partial_\|$, and $X,Y=\delta,v$ or $ \Phi$. The corresponding term in the kernel is formed as follows:
\bea
&\Big\{& {1\over2}\big({\rm i}\, k_1\big)^n \big({\rm i}\, k_2\big)^m  ~\text{for D\,$=\partial_\|$ OR }~{1\over2}\big({\rm i}\, \k_1\cdot {\rm i}\, \k_2\big)^n~\text{ for D\,$ =\partial_i$, $m=n$} \nonumber\\
&\times& \big[k_1^{-2} \text{~if $X$ is $v$ or $\Phi$}\big] \times \big[k_2^{-2} \text{~if $Y$ is $v$ or $\Phi$}\big] \nonumber\\
&\times&  \big[\text{a factor of }~\mu_1 \text{~for each $\partial_\|$ acting on $X$} \big] \times \big[\text{a factor of}~\mu_2 \text{~for each $\partial_\|$ acting on $Y$} \big] \nonumber\\
&\times& \big[\text{a factor of }~f\HH\text{~ for each~} v\big]\nonumber\\
&\times& \big[\text{a factor of }~-\frac{3}{2} \Omega_m \HH^2\text{~ for each~} \Phi\big] \Big\}\nonumber\\
&+&~\Big\{1\leftrightarrow 2\Big\}
\eea


\newpage
\section{The coefficients in the GR kernel ${{\cal K}^{(2)}_{\rm GR}}$}

{The coefficients $\Gamma_I(z)$ in \eqref{e8} follow from \eqref{eq:SecondorderNewtonian}--\eqref{bce}, using Table~\ref{table1}.}

Evolution bias $b_e$ and magnification bias ${\cal Q}$ make the $\Gamma_I$ much more complicated than for the case
$b_e=0={\cal Q}$, which is considered in \cite{Umeh:2016nuh}.
 (Note that when ${\cal Q}=0$, all the terms with $\partial/\partial\ln\bar L$ vanish.)

\begin{eqnarray} 
\frac{\Gamma_{1}}{\mathcal{H}^{4}} &=& \frac{9}{4}\Omega_{m}^{2}\Bigg[{-3} +2f\bigg({2}-2b_{e}+{4}\Q+\frac{4(1-\Q)}{\chi\HH} +\frac{2\HH'}{\HH^{2}}\bigg) -\frac{2f'}{\HH} +b_{e}^{2}+ 6b_{e}-8b_{e}\Q+4\Q+16\Q^{2} -16\frac{\p \Q}{\p \ln\bar{L}} \nonumber\\ 
&&{} -8\frac{\Q'}{\HH} + \frac{b_{e}'}{\HH}+\frac{2}{\chi^{2}\HH^{2}}\bigg(1-\Q+2\Q^{2}-2\frac{\p \Q}{\p \ln{\bar{L}}}\bigg)- \frac{2}{\chi\HH}\bigg(4+2b_{e}-2b_{e}\Q-4\Q+8\Q^{2}-\frac{3\HH'}{\HH^{2}}(1-\Q) \nonumber \\ 
&&{}- 8\frac{\p \Q}{\p \ln{\bar{L}}} - 2\frac{\Q'}{\HH}\bigg) + \frac{\HH'}{\HH^{2}}\bigg(-8-2b_{e}+{8}\Q+\frac{3\HH'}{\HH^{2}}\bigg) - \frac{\HH''}{\HH^{3}}\Bigg] \nonumber \\
&&{} +3\Omega_{m}f\Bigg[6-f(3-b_{e})+b_{e}\bigg(3+\frac{2(1-\Q)}{\chi\HH}\bigg)-\frac{b_{e}'}{\HH}-b_{e}^{2}+4b_{e}\Q-12\Q-\frac{6(1-\Q)}{\chi\HH}+2\bigg(2-\frac{1}{\chi\HH}\bigg)\frac{\Q'}{\HH}\Bigg] \nonumber \\
&&{}+f^{2}\Bigg[12-7b_{e}+b_{e}^{2}+\frac{b_{e}'}{\HH}+(b_{e}-3)\frac{\HH'}{\HH^{2}}\Bigg] \\
\frac{\Gamma_{2}}{\mathcal{H}^{3}} &=& \frac{9}{4}\Omega_{m}^{2}(f-2+2\Q) + \frac{3}{2}\Omega_{m}f\Bigg[-2 -f\bigg(-3+f+2b_{e}-3\Q-\frac{4(1-\Q)}{\chi\HH}-\frac{2\HH'}{\HH^{2}}\bigg)-\frac{f'}{\HH}+3b_{e}+b_{e}^{2}-6b_{e}\Q+ {4}\Q\nonumber \\
&&{}+8\Q^{2}-8\frac{\p \Q}{\p \ln{\bar{L}}} -6\frac{\Q'}{\HH} +\frac{b_{e}'}{\HH} +\frac{2}{\chi^{2}\HH^{2}}\bigg(1-\Q+2\Q^{2}-2\frac{\p \Q}{\p \ln{\bar{L}}}\bigg) + \frac{2}{\chi \HH}\bigg(-1 -2b_{e}+2b_{e}\Q+\Q-6\Q^{2}  \nonumber \\
&&{} +\frac{3\HH'}{\HH^{2}}(1-\Q) +6\frac{\p \Q}{\p \ln{\bar{L}}} + 2\frac{\Q'}{\HH}\bigg) -\frac{\HH'}{\HH^{2}}\bigg(3+2b_{e}-6\Q-\frac{3\HH'}{\HH^{2}}\bigg) - \frac{\HH''}{\HH^{3}}\Bigg] + f^{2}\Bigg[-3+2b_{e}\bigg(2+\frac{(1-\Q)}{\chi\HH}\bigg)\nonumber \\
&&{}-b_{e}^{2}+2b_{e}\Q -6\Q-\frac{b_{e}'}{\HH}-\frac{6(1-\Q)}{\chi\HH}+2\bigg(1-\frac{1}{\chi\HH}\bigg)\frac{\Q'}{\HH}\Bigg] \\
\frac{\Gamma_{3}}{\HH^{2}} &=& \frac{3}{2}\Omega_{m}\Bigg[2-2f+b_{e}-4\Q-\frac{2(1-\Q)}{\chi\HH}-\frac{\HH'}{\HH^{2}}\Bigg] \\
\frac{\Gamma_{4}}{\HH^{2}} &=& f(3-b_{e}) \\
\frac{\Gamma_{5}}{\HH^{2}} &=& 3\Omega_{m}f(2-f-2\Q) + f^{2}\Bigg[4+b_{e}-b_{e}^{2}+4b_{e}\Q-{6}\Q-4\Q^{2}+4\frac{\p \Q}{\p \ln{\bar{L}}} + 4\frac{\Q'}{\HH} - \frac{b_{e}'}{\HH}  \nonumber \\
&&{} - \frac{2}{\chi^{2}\HH^{2}}\bigg(1-\Q+2\Q^{2}-2\frac{\p \Q}{\p \ln{\bar{L}}}\bigg) - \frac{2}{\chi\HH}\bigg(3-2b_{e}+2b_{e}\Q-\Q-4\Q^{2}+\frac{3\HH'}{\HH^{2}}(1-\Q) + 4\frac{\p \Q}{\p \ln{\bar{L}}} + 2\frac{\Q'}{\HH}\bigg) \nonumber \\
&&{} - \frac{\HH'}{\HH^{2}}\bigg(3-2b_{e}+{4}\Q+\frac{3\HH'}{\HH^{2}}\bigg) + \frac{\HH''}{\HH^{3}}\Bigg]  \\
\frac{\Gamma_{6}}{\HH^{2}} &=& 3\Omega_{m}f - f^{2}\Bigg[-1+b_{e}-2\Q- \frac{2(1+\Q)}{\chi\HH}-\frac{\HH'}{\HH^{2}}\Bigg] \\
\frac{\Gamma_{7}}{\HH^{2}} &=& \frac{3}{2}\Omega_{m}\Bigg[b_{1}\bigg(2+b_{e}-4\Q-\frac{2(1-\Q)}{\chi\HH} -\frac{\HH'}{\HH^{2}}\bigg) + \frac{b_{1}'}{\HH} + 2\bigg(2-\frac{1}{\chi\HH}\bigg)\frac{\p b_{1}}{\p \ln{\bar{L}}}\Bigg] - f\Bigg[b_{1}(f-3+b_{e}) + \frac{b_{1}'}{\HH}\Bigg]  \\
\frac{\Gamma_{8}}{\HH^{2}} &=& \frac{9}{4}\Omega_{m}^{2} + \frac{3}{2}\Omega_{m}f\Bigg[{1}-2f+2b_{e}-{6}\Q-\frac{4(1-\Q)}{\chi\HH}-\frac{3\HH'}{\HH^{2}}\Bigg] + f^{2}(3-b_{e}) \\ \nonumber \\ \nonumber
\end{eqnarray}
\begin{eqnarray}
\nonumber \\
\frac{\Gamma_{9}}{\HH} &=& -\frac{3}{2}\Omega_{m}b_{1} \\
\frac{\Gamma_{10}}{\HH} &=& 2f^{2}\\
\frac{\Gamma_{11}}{\HH} &=& f\Bigg[b_{1}\bigg(f+b_{e}-2\Q-\frac{2(1-\Q)}{\chi\HH}-\frac{\HH'}{\HH^{2}}\bigg) + \frac{b_{1}'}{\HH} + 2\bigg(1-\frac{1}{\chi\HH}\bigg)\frac{\p b_{1}}{\p \ln \bar{L}}\Bigg] \\
\frac{\Gamma_{12}}{\HH} &=& -\frac{3}{2}\Omega_{m}f \\
\frac{\Gamma_{13}}{\HH} &=& \frac{3}{2}\Omega_{m}f -f^{2}\Bigg[3-2b_{e}+{4}\Q+\frac{4(1-\Q)}{\chi\HH}+\frac{3\HH'}{\HH^{2}}\Bigg] \\
\frac{\Gamma_{14}}{\HH} &=& f\Bigg[b_{e}-2Q-\frac{2(1-\Q)}{\chi\HH}-\frac{\HH'}{\HH^{2}}\Bigg]
\end{eqnarray}

\newpage

\bibliographystyle{JHEP}
\bibliography{PNG_Ref}

\end{document}